\documentclass{pasj00}
\draft

\begin{document}
\SetRunningHead{Morii et al.}{AXP 1E 1841$-$045}
\Received{2010/07/07}
\Accepted{2010/07/07}

\title{Suzaku Observation of the Anomalous X-ray Pulsar 1E 1841$-$045}


 \author{%
   Mikio \textsc{Morii}\altaffilmark{1,2},
   Shunji \textsc{Kitamoto}\altaffilmark{2},
   Noriaki \textsc{Shibazaki}\altaffilmark{2},
   Nobuyuki \textsc{Kawai}\altaffilmark{1}, \\
   Makoto \textsc{Arimoto}\altaffilmark{1}, 
   Masaru \textsc{Ueno}\altaffilmark{1},
   Takayoshi \textsc{Kohmura}\altaffilmark{3},
   Yukikatsu \textsc{Terada}\altaffilmark{4}, \\
   Shigeo \textsc{Yamauchi}\altaffilmark{5},
   and
   Hiromitsu \textsc{Takahashi}\altaffilmark{6}
   }
 \altaffiltext{1}{Department of Physics, Tokyo Institute of Technology,
   Ookayama 2-12-1, Meguro-ku, \\
   Tokyo 152-8551, Japan}
 \email{morii.m.ab@m.titech.ac.jp, mmorii@rikkyo.ac.jp}
 \altaffiltext{2}{Department of Physics, Rikkyo University, Nishi-ikebukuro 3-34-1,
Toshima-ku, Tokyo 171-8501, Japan}
 \altaffiltext{3}{Physics Department, Kogakuin University 2665-1 Nakano-cho,
Hachioji, Tokyo 192-0015, Japan}
 \altaffiltext{4}{Cosmic Radiation Laboratory,
 Institute of Physical and Chemical Research, Wako, Saitama 351-0198, Japan}
 \altaffiltext{5}{Department of Physics, Faculty of Science, Nara Women's University
Kitauoyanishi-machi, \\
Nara 630-8506, Japan}
 \altaffiltext{6}{Department of Physical Science, School of Science, Hiroshima University 1-3-1 Kagamiyama, \\
 Higashi-Hiroshima, Hiroshima 739-8526, Japan}


\maketitle

\begin{abstract}
We report the results of a Suzaku observation of the anomalous X-ray pulsar
(AXP) 1E 1841$-$045 at a center of the supernova remnant Kes 73.
We confirmed that the energy-dependent spectral models obtained by the previous separate observations
were also satisfied over a wide energy range from 0.4 to $\sim$70 keV, simultaneously. 
Here, the models below $\sim$10 keV were a combination of blackbody (BB) and power-law (PL) functions or of two BBs with different temperatures at 0.6 -- 7.0 keV \citep{Morii et al. 2003},
and that above $\sim$20 keV was a PL function \citep{Kuiper Hermsen Mendez 2004}.
The combination ${\rm BB} + {\rm PL} + {\rm PL}$ was found to best represent the 
phase-averaged spectrum.
Phase-resolved spectroscopy indicated the existence of two emission regions,
one with a thermal and the other with a non-thermal nature.
The combination BB + BB + PL was also found to represent the phase-averaged spectrum well.
However, we found that this model is physically unacceptable
due to an excessively large area of the emission region of the blackbody.
Nonetheless, we found that the temperatures and radii of the two blackbody components
showed moderate correlations in the phase-resolved spectra.
The fact that the same correlations have been observed between the
phase-averaged spectra of various magnetars \citep{Nakagawa et al. 2009}
suggests that a self-similar function can approximate
the intrinsic energy spectra of magnetars below $\sim10$ keV. 

\end{abstract}

\section{Introduction} \label{sec: Introduction}
Anomalous X-ray pulsars (AXPs) are thought to be magnetars,
which are strongly magnetized 
neutron stars with emissions
powered by dissipation of magnetic energy
 (see \citet{Woods Thompson 2006} for a recent review).
So far, about 10 AXPs have been reported as well-confirmed objects or candidates.
They are distributed along the Galactic plane and within the Small Magellanic Cloud.
They are characterized as follows:
(1) The spin periods ($P$) are concentrated within the narrow range of 2 -- 12 s.
    In addition, their spin-down rates ($\dot{P}$) are
    large ($5 \times 10^{-13}$ -- $1 \times 10^{-10}$ s s$^{-1}$).
    Both of these facts
    imply that they have strong dipole magnetic fields 
    of $10^{14}$ -- $10^{15}$ G at their surfaces.
(2) Their X-ray luminosities ($L_{\rm X} \sim 10^{34}$--$10^{36}$ ergs s$^{-1}$) exceed
    the rate of spin-down energy loss of neutron stars
    ($\dot{E} = 4 \pi^2 I \dot{P}/ P^3 \sim 10^{32.6}$ ergs s$^{-1}$,
    where $I \simeq 10^{45}$ g cm$^2$ is the momentum of inertia of a neutron star.).
    This fact suggests that AXPs are not rotation-powered pulsars.
(3) Timing characteristics peculiar to accretion-powered pulsars, such as
    a large amount of timing noise and persistent spin-up periods
    have not been observed to date.
(4) Whereas persistent radio emissions have not been detected,
    transient pulsed radio emissions were found
    for transient AXPs, XTE J1810$-$197 \citep{Camilo et al. 2006}
    and 1E 1547.0$-$5408 \citep{Camilo et al. 2007}.
(5) Some AXPs produced spiky short bursts with durations $\Delta t \sim 0.1$ s,
    similar to those observed in soft gamma repeaters (SGRs), which are also magnetar candidates.
(6) They have peculiar energy spectra which will be described below.

Energy spectra of AXPs below $\sim 10$ keV can empirically be modeled well by
either a combination of a blackbody and a power-law function or two blackbody functions.
Fitting the spectra using the former model yields a
blackbody temperatures of about $0.4$ keV
and the photon indices are within the range of $2.0 - 4.6$ \citep{Woods Thompson 2006}.
Recently, using the latter model, \citet{Nakagawa et al. 2009} reported that
there are correlations between the temperatures and radii of the two blackbodies.
On the other hand, non-thermal,
hard X-ray emissions with large pulse fractions
were discovered from some AXPs above $\sim 10$ keV
\citep{Kuiper et al. 2006}.
Their total and pulsed spectra were found to have power-law shapes.
The photon indices of the total spectra were in a range from 1.0 to 1.4,
and those of the pulsed spectra in a range from $-1.0$ to 1.0.
These emissions were detected up to $\sim 220$ keV
\citep{Kuiper et al. 2006, den Hartog et al. 2008, den Hartog Kuiper Hermsen 2008}.
Some theoretical models of the hard X-ray emission were proposed, based on thermal bremsstrahlung
\citep{Thompson Beloborodov 2005, Beloborodov Thompson 2007},
synchrotron radiation
\citep{Thompson Beloborodov 2005, Beloborodov Thompson 2007},
and emission caused by QED (quantum electrodynamics) effects
\citep{Heyl Hernquist 2005}.
However, the emission mechanism is still not clearly understood.

One AXP candidate is 1E 1841$-$045
\citep{Hellier 1994, Mereghetti Stella 1995, van-Paradijs Taam van-den-Heuvel 1995}.
It is located at the center of the supernova remnant (SNR) Kes 73
(G 27.4+0.0; \cite{Kriss et al. 1985}) with a diameter of about \timeform{4'}
\citep{Morii et al. 2003}.
The kinematic distance to the SNR was originally estimated to be between
6 and 7.5 kpc \citep{Sanbonmatsu Helfand 1992}.
Recently, this distance was revised to be
between 7.5 and $\sim 9.8$ kpc by \citet{Tian Leahy 2007}.
The pulse period of 1E 1841$-$045 was determined to be 11.8 s \citep{Vasisht Gotthelf 1997}.
Its activity has been continuously monitored by the Rossi X-ray Timing Explorer (RXTE) \citep{Dib Kaspi Gavriil 2008},
and although several glitches occurred during the monitoring period, there is no evidence
for any pulsed flux change in the 2$-$10 keV band. 
Using Chandra data in the 0.6 -- 7.0 keV region, \citet{Morii et al. 2003} reported that the spectrum could be modeled well
using a model consisting of a blackbody ($kT = 0.44 \pm 0.02$ keV)
and a power-law function with the hardest photon index yet measured for
AXPs ($\Gamma = 2.0 \pm 0.3$).
In their analysis, they showed that the two-blackbody model
($kT = 0.47 \pm 0.02$ keV and $kT = 1.5^{+0.2}_{-0.1}$ keV) also fitted well.
In addition, they showed that there were two emission regions
and the emissions appeared as two peaks in the pulse profile.
The peaks were interpreted as blackbody emission from a hot spot
and non-thermal emission from another region,
although blackbody emissions from two hot spots were
also a possible explanation.
\citet{Kuiper Hermsen Mendez 2004} discovered the very hard
(photon index: $0.94 \pm 0.16$) pulsed X-ray emission
up to $\sim 150$ keV, using the RXTE Proportional Counter Array (PCA; 1.8 -- 23.8 keV) and
the High Energy X-ray Timing Experiment (HEXTE; 15 -- 250 keV). 
The discovery of the hard X-ray component compels a reconsideration of
the spectral models below $\sim 10$ keV, because this component
is thought to account for a substantial fraction in an energy range below $\sim 10$ keV.

In this paper, we report on results obtained from a Suzaku observation of AXP 1E 1841$-$045.
Thanks to the wide energy coverage of Suzaku,
we obtained a continuous spectrum from 0.4 to $\sim 70$ keV.
Spectroscopic data over such a wide energy band offers advantages in
decomposing a spectrum into its constituent components. 

\section{Observation}
1E 1841$-$045 and Kes 73 were observed
by the fifth Japanese X-ray satellite Suzaku on April 19--22, 2006,
as a target of AO-1.
%
Suzaku \citep{Mitsuda et al. 2007} has two types of X-ray detectors in operation:
the X-ray Imaging Spectrometer (XIS; \cite{Koyama et al. 2007}) and
the Hard X-ray Detector (HXD; \cite{Takahashi et al. 2007, Kokubun et al. 2007}).
Four XISs are mounted at the foci of four X-ray telescopes
(XRT; \cite{Serlemitsos et al. 2007}) with a moderate angular resolution of \timeform{2}
(half power diameter; HPD).
Three of the XISs (XIS0, XIS2, and XIS3) are front-illuminated (FI) CCDs,
sensitive in the energy range of 0.4--12 keV,
and the remaining one (XIS1) is a back-illuminated (BI) CCD,
sensitive in the energy range of 0.2--12 keV.
We chose the ``1/8 window mode'' of the XISs
to obtain a time resolution of 1 s, which corresponds to 0.08 times the pulsation period
of this AXP.
This mode covered a field of view (FoV)
of \timeform{17.8'} $\times$ \timeform{2.2'} ($1024 \times 128$ pixels).

The HXD is a non-imaging, collimated hard X-ray detector sensitive in
the energy range of 10--600 keV.
It is composed of PIN photodiodes and GSO scintillators
mounted in a well of collimator shields,
which cover the energy regions 10$-$70 keV and 50$-$600 keV,
with a field of view of \timeform{34'} $\times$ \timeform{34'} ($\lesssim$ 100 keV)
and \timeform{4.5D} $\times$ \timeform{4.5D} ($\gtrsim$ 100 keV), respectively.
The time resolution during our observations was 61 $\mu$s.

Our observations were carried out at the position
(RA, Dec) $=$ (\timeform{18h41m15.5s}, \timeform{-4D51'24.5''})
(J2000.0) at the HXD nominal pointing. 

\section{Analysis}\label{sec: analysis}

We followed the procedure outlined in the ``seven steps`` manuals for XIS and HXD analyses
\footnote{http://www.astro.isas.jaxa.jp/suzaku/analysis/7step\_XIS\_20061218.txt}$^,\!\!$
\footnote{http://www.astro.isas.jaxa.jp/suzaku/analysis/7step\_HXD\_20070305.txt}.
We used the cleaned events generated by the standard pipeline processing of rev-1.2.

For the XIS analysis, we removed the timeslice 13:05 -- 14:42 (UTC) on April 20, 2006,
because a temperature anomaly of XIS2 was reported
\footnote{http://www.astro.isas.jaxa.jp/suzaku/log/xis/}.
We selected as source and background regions a concentric circle 
and annulus whose central positions were at the peak position of the AXP.
The radius of the source circle was set to \timeform{4.33'},
and the radii of the inner and outer boundaries of
the background annulus were set to \timeform{4.33'} and \timeform{6.00'}, respectively.
We then set the source and background regions as
the intersections of these concentric regions and the 1/8 window regions
of the XISs.
The net exposure times for the XISs were 95.3 ks. 

For the PIN and GSO analyses
we used a background file made by the LCFIT (bgd\_d) method of version 1.2 v.0611.
This observation was performed in the period from March 14 to May 13, 2006,
when the HXD was operated with a lower discriminator level for the PSD cut.
This caused an increase in the GSO events and
saturation of the internal data transfer sometimes occurred,
especially during periods of high count rates due to background particles. 
Since the uncertainty in the source flux was higher during these periods,
we applied an additional GTI (good time interval) to remove these periods according to the recommendations of the detector team
\footnote{http://www.astro.isas.jaxa.jp/suzaku/analysis/hxd/hxdgti/}
in addition to the normal GTI made by the event and background files.
The net exposures for the PIN and GSO were 57.8 ks.

We checked the light curves of all XISs and the PIN binned at 300, 10 and 1 s, and
found no notable variation due to background changes, instrument trouble 
or AXP activity.

\section{Timing Analysis} \label{sec: timing analysis}

It is known that the time tagged for each XIS frame 
in the 1/8 window mode was 7 s
earlier than that of the HXD \footnote{http://www.astro.isas.ac.jp/suzaku/analysis/xis/timing/},
whereas the time tagging of the HXD is
well calibrated by comparison with other satellites, such as RXTE,
INTEGRAL, and Swift \citep{Terada et al. 2008}.
We corrected the XIS frame times by adding 7 s to the reference time "MJDREFF".
We corrected photon arrival times into those at the solar barycenter by
using \texttt{aebarycen} (ver. 2006-08-02).
We constructed the overall light curve by summing the light curves of all the XISs 
for the source region with 1 s binning in an energy range of 0.4 -- 10.0 keV.
We searched for periodicity in this light curve using \texttt{xronos/powspec}
and determined the pulse period more precisely using \texttt{xronos/efsearch}.
The pulse period obtained in this manner was $11.7830 \pm 0.0002$ s
at the epoch of 53845.7951888 (MJD),
which is consistent with the phase coherent ephemeris obtained
by \citet{Dib Kaspi Gavriil 2008}.
We also searched for periodicity in the PIN light curve
in the energy range of 12.0 -- 50.0 keV using the $Z_1^2$ method \citep{Buccheri+1983}.
We detected the signal significantly in 
the null hypothesis probability of $1.2\times10^{-5}$.
For the GSO light curves (40 -- 60, 50 -- 100 and 100 -- 600 keV)
we could not find a signal
corresponding to the pulse period of the AXP.

\begin{figure}
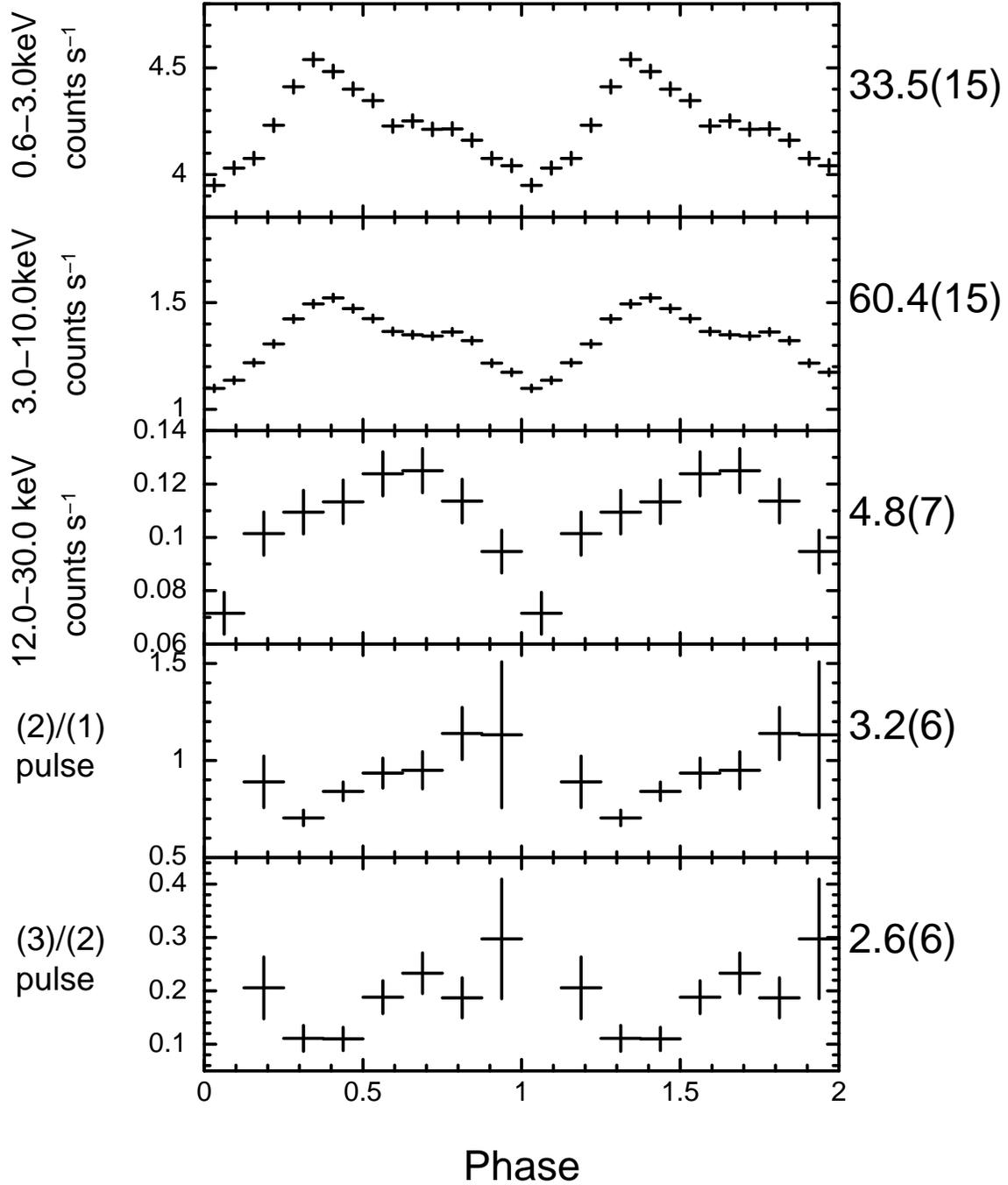

  \begin{center}
    \FigureFile(150mm,150mm){figure1.ps}
  \end{center}
  \caption{Folded pulse profiles and hardness ratios.
The top two panels show pulse profiles made by the sum of all XISs
in the energy ranges of 0.6 -- 3.0 keV and 3.0 -- 10.0 keV.
The third panel shows a pulse profile for the PIN in an
energy range of 12 -- 30 keV.
The vertical axes have units of count rate (counts s$^{-1}$)
after the corresponding background-subtractions
(section \ref{sec: timing analysis}).
The fourth and fifth panels show the hardness ratios of
the pulsed component, which are
the ratios of the photon count rates
in the second to the first panel
and the third to the second, respectively,
after subtracting the photon count rates
at the pulse minimum phases (phase = 0.0 -- 0.125).
The horizontal axes show the pulse phase up to
2 periods. The vertical error bars in all panels represent $1\sigma$ levels.
At the right of each panel the reduced $\chi^2$ and
the degree of freedom are shown
for the case of fitting the profile to a constant model.
}\label{fig: prof+hr.ps}
\end{figure}

We made pulse profiles for three energy ranges
(0.6 -- 3.0, 3.0 -- 10.0 and 12.0 -- 30.0 keV)
by folding the light curves with the pulse period obtained
at the same epoch (figure \ref{fig: prof+hr.ps}).
Here, for the XIS light curves (0.6 -- 3.0 and 3.0 -- 10.0 keV),
non-X-ray and X-ray backgrounds were subtracted
by selecting the background region as described in section \ref{sec: analysis}.
For the PIN light curve (12.0 -- 30.0 keV)
only the non-X-ray background ($0.337\pm0.010$(syst) counts s$^{-1}$
with a systematic uncertainty of 3\%.
\footnote{Suzaku-memo 2006-42, 2006-43: http://www.astro.isas.jaxa.jp/suzaku/doc/suzakumemo/\label{fn: suzaku-memo}})
was subtracted and the X-ray background was retained.
The pulse fraction ($PF$), defined as
$PF = \left[\sum (R_i - R_{\rm min})\right]/ \sum R_i$,
for these bands was $6.6 \pm 0.7$(stat)\%,
$17.2 \pm 1.1$(stat)\%
and $32.9 \pm 7.0$(stat)$\pm 8.9$(syst)\%, respectively.
Here, $R_i$ and $R_{\rm min}$ denote the count rate
in the $i$-th phase bin and the minimum
count rate among all the phase bins, respectively.
The errors denoted by ``(stat)'' and ``(syst)''
represent a statistical error of 1$\sigma$, and
a systematic error due to the 3\% uncertainty in the non-X-ray background
of the PIN \footnotemark[\ref{fn: suzaku-memo}].
We will use this convention from this point onwards.
In the PIN energy range (12.0 -- 30.0 keV)
the pulsed fraction increased to $60.8 \pm 13.1$(stat)$\pm 16.8$(syst)\%
after subtracting both the non-X-ray and X-ray backgrounds.
Here, we estimated the count rate for the X-ray background
to be 0.0489 counts s$^{-1}$ (subsection \ref{subsection: Preparation}).
Although this is larger than the value of $\sim25$\% at 20 keV reported by \citet{Kuiper et al. 2006},
the difference is not significant
due to the large uncertainty in the pulsed fraction for the PIN.

We also made plots of the hardness ratios
of the second pulse profile to the first, and
of the third to the second.
To eliminate the constant component
the count rate at the pulse minimum phase (0.0 -- 0.125)
was subtracted before the division.
It can be seen that 
the spectrum at the second peak of the pulse profile
was harder than that of the first,
which is consistent with
the result obtained by the previous Chandra observation
(e.g. \cite{Morii et al. 2003}).

\section{Energy Spectrum}

\subsection{Spectroscopy of the SNR Kes 73 using the Chandra Data}

\begin{figure}
  \begin{center}
    \FigureFile(120mm,120mm){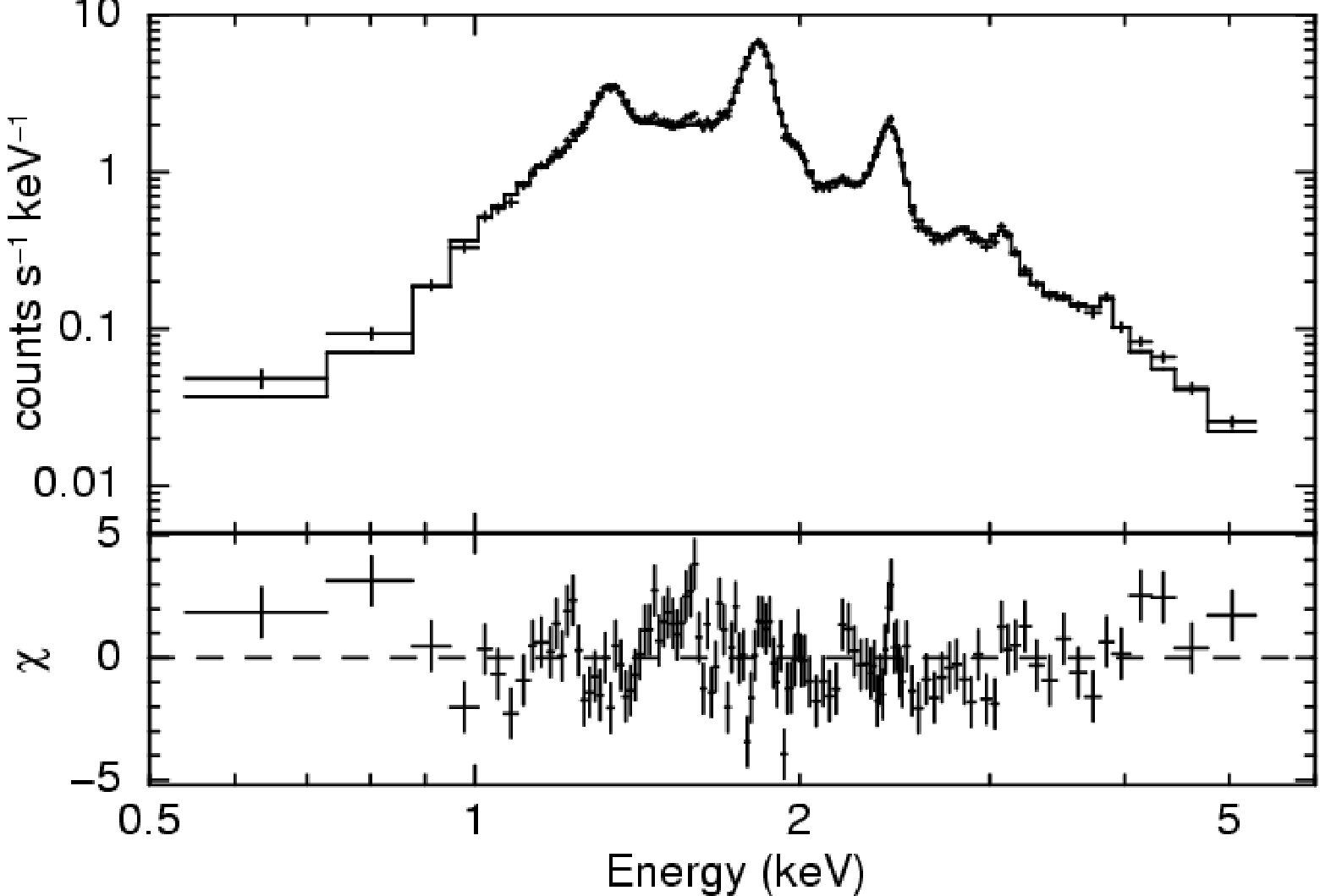}
  \end{center}
  \caption{Spectrum of Kes 73 taken by Chandra,
	 fitted by the \texttt{vsedov} model with the optimum parameters.
	 The vertical bars represent $1\sigma$ errors.
	 }\label{fig: wabs+vsedov_07021100.ps}
\end{figure}

Due to the point spread function (PSF) of the XRTs (HPD = \timeform{2'})
the spectrum of the AXP was contaminated by the surrounding SNR Kes 73.
Therefore, we must determine the spectral model of the SNR
before performing spectral analyses with the Suzaku data.
For this purpose, we analyzed archival data of the Chandra observation
taken with the AXAF CCD Imaging Spectrometer (ACIS) in the TE mode (Observation ID: 729).
We extracted the pure SNR spectrum by spatially excluding the AXP and the charge trailing regions.
We then attempted to fit the entire SNR spectrum using the \texttt{vnei}, \texttt{vpshock}, and \texttt{vsedov}
models \citep{Borkowski et al. 2001}.
It was found that the entire SNR spectrum could be modeled by the \texttt{vsedov} model
(see figure \ref{fig: wabs+vsedov_07021100.ps}).
Although even the best model was statistically unacceptable,
globally, the fit appeared good and residuals were present only at the emission lines.
Therefore, we believe that such local residuals in the SNR model
would not affect the interpretation of the Suzaku spectroscopic data for the AXP spectrum,
because the AXP spectrum is composed only of continuum components.

Moreover, the spreading due to the wide angular response of the XRTs would effectively homogenize 
spatial variation in the SNR spectrum.
Therefore, we can assume that the spectrum of the SNR is spatially uniform.
Consequently, in the following section
we will use the SNR spectral model obtained here as the SNR component.
Further analysis was carried out on the SNR data,
but the details are beyond the scope of this paper, and will be published at a later date.

\subsection{Preparation of the Suzaku Data: Estimation of Backgrounds and Calibrations}
\label{subsection: Preparation}

We analyzed the XISs and PIN source spectra
after subtracting the corresponding backgrounds.
Here, the source and background spectra of XISs were
extracted from the regions shown in section \ref{sec: analysis}
and the non-X-ray background 
was subtracted in the case of the PINs.
We made the response files
(RMF; Redistribution Matrix File and ARF; Ancillary Response File) of the XISs
by using \texttt{xisrmfgen} and \texttt{xissimarfgen}.
For the response of the PIN, we used that
for a point source at the nominal position of the HXD 
(\texttt{ae\_hxd\_pinhxnom\_20060814.rsp}).
For the spectral fits we used \texttt{xspec11} in \texttt{HEADAS v6.3.2}.
We rebinned the PI spectra of the XISs
so that at least $N_{\rm min}$ counts could be contained in each bin
by using \texttt{grppha}.
Here, we set the counts $N_{\rm min}$ to
50, 100, 500, 100, and 50 
for energy intervals of
less than 0.9, 0.9 -- 1.2, 1.2 -- 3.0, 3.0 -- 5.0,
and more than 5.0 keV, respectively.
We used $N_{\rm min} = 800$ counts for the PIN spectrum.
We followed the next steps for the spectral fitting.

(1) To fix the gain uncertainty of the XISs
we fitted only the XISs data by a model,
$c_{\rm det} \times \left( c_{\rm SNR} \times {\rm SNR} + {\rm AXP} \right)$,
while the gain parameters (energy scales and offsets) were allowed to vary.
Here, ${\rm SNR}$ and ${\rm AXP}$ are self-explanatory spectral components.
The parameter $c_{\rm SNR}$ denotes the normalization of the SNR component.
It was used to compensate for the discrepancy in the SNR flux between
this observation and the Chandra observation,
which was caused by the small coverage area of the SNR region in the 1/8 window mode
of the XISs.
The parameters $c_{\rm det}$ (det = XIS0, XIS1, XIS2, and XIS3) represent
normalizations of the detectors.
They were used to compensate for the relative flux uncertainty among the XISs,
fixing $c_{\rm XIS0}$ to be 1.
This flux uncertainty was caused by the systematic uncertainty of the ARF
generated by \texttt{xissimarfgen},
which has a systematic uncertainty 
for small regions such as those used in this analysis (section \ref{sec: analysis}).

(2) Fixing the gain parameters determined using the first step,
we fitted the XISs and PIN data simultaneously using a model,
$c_{\rm det} \times \left( c_{\rm SNR} \times {\rm SNR} + {\rm AXP}
 + {\rm CXB} + {\rm GRXE} \right)$.
Here, ${\rm CXB}$ and ${\rm GRXE}$ denote the spectral components
from the cosmic X-ray background
and the Galactic ridge X-ray emission.
The ${\rm CXB} + {\rm GRXE}$ component was taken into consideration only for the HXD/PIN,
because in the case of the XISs this component could be removed by the background subtraction
shown in section \ref{sec: analysis}.

The ${\rm CXB} + {\rm GRXE}$ component was estimated by using additional Suzaku data
from the GRXE observation at the field ($l =$ \timeform{28.5D}, $b =$ \timeform{-0.2D}).
Fortunately, this field is very close to the pointing field of the AXP,
($l =$ \timeform{27.5D}, $b =$ \timeform{0.0D}),
so this estimation is expected to produce a reasonable result.
The spectrum of the PIN at ($l =$ \timeform{28.5D}, $b =$ \timeform{-0.2D}) was fitted by
a power-law function (``pegpwrlw'' model)
using the response file for a point source
at the nominal position of the HXD (\texttt{ae\_hxd\_pinhxnom\_20060814\_w23.rsp}).
This fitting resulted in a normalization of
$17.5^{+2.0}_{-1.8}$ (stat) $\times 10^{-12}$ erg/cm$^2$/s 
in the energy range 10 -- 50 keV
and a photon index of $2.64^{+0.43}_{-0.39}$(stat).
The flux of the GRXE component at the AXP field ($l =$ \timeform{27.5D}, $b =$ \timeform{0.0D})
would be slightly different from that
at the GRXE field ($l =$ \timeform{28.5D}, $b =$ \timeform{-0.2D}),
because the GRXE intensity depends on 
the galactic latitude (e.g., \cite{Yamauchi Koyama 1993, Kaneda et al. 1997}).
We assumed that the spatial distribution of the GRXE is modeled by
$F(|b|) = F_0 \exp(- |b|/h)$ with a scale height of $h =$ \timeform{0.5D} and
estimated the difference in photon counts in the FoVs of the PIN for both pointings
through simulation.
As a result, we found that the GRXE flux at the AXP field was about 10\% larger
than that at the GRXE field.
The count rate of 10\% of the GRXE + CXB amounts to only 1\%
of the non-X-ray background of the PIN in the energy range of 12 -- 50 keV.
Therefore, we neglected the difference of the GRXE count rates between
the GRXE and the AXP fields.
Consequently, we decided to use the best fitting model
at the field ($l =$ \timeform{28.5D}, $b =$ \timeform{-0.2D})
as the CXB + GRXE component at the AXP field in subsequent analysis.

(3) Because this observation was performed in the 1/8 window mode of the XISs,
a substantial amount of photons from the point source were missing from the FoV.
Due to the systematic uncertainty of the PSFs of the XRTs,
a systematic discrepancy between the normalization of the XISs
and that of the PIN was inevitable.
Therefore, we calibrated the cross normalization between the XISs and PIN, using
the archival data of the point source, MCG$-$5$-$23$-$16 (data ID: 700002010),
which was observed at the HXD nominal pointing and
taken in the full window mode of the XISs.
We produced the XIS spectra by extracting photons from the region in a manner similar to that described in section \ref{sec: analysis}, setting the distance
from the peak position of the point source to
a long side of the rectangular FoV of the 1/8 window region
to be equal to that for the case of the AXP.
We obtained a relative normalization between XIS0 and the PIN of 1.0890,
which was used in subsequent analysis.

\subsection{Phase-averaged Spectroscopy}\label{subsection: Phase-averaged Spectroscopy}

We first tried applying a power-law function (${\rm PL}$)
\footnote{We used ``pegpwrlw'' model instead of ``powerlaw'' model in XSPEC
to reduce the off-axis elements of the error matrix used to evaluate the uncertainties
of the parameters.
In the ``pegpwrlw'' model we set the energy range 1 - 50 keV over which the flux is integrated.},
blackbody radiation (${\rm BB}$), and thermal bremsstrahlung (${\rm TB}$) to the ${\rm AXP}$ component.
However, these were found to be statistically unacceptable.
Among the pairs
(${\rm PL} + {\rm PL}$, ${\rm BB} + {\rm BB}$, ${\rm TB} + {\rm TB}$,
${\rm PL} + {\rm BB}$, ${\rm PL} + {\rm TB}$, ${\rm BB} + {\rm TB}$),
all except the ${\rm TB} + {\rm TB}$ showed excesses at a higher energy band
or residuals, suggesting it was necessary to add an additional component
at the higher energy range.

\begin{figure}
  \begin{center}

    \FigureFile(150mm,150mm){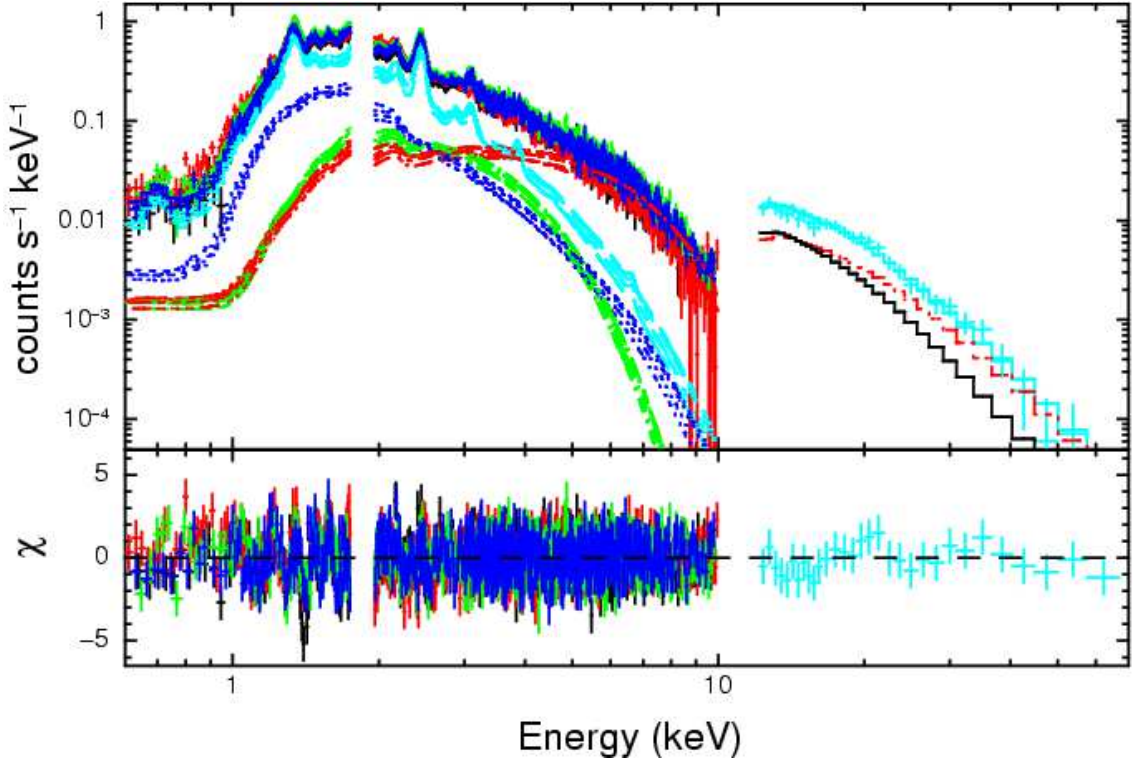}
  \end{center}
  \caption{Phase-averaged spectrum fitted with the BB + PL + PL model
after background subtraction
(section \ref{sec: analysis} and subsection \ref{subsection: Preparation}).
The data points from XIS0, XIS1, XIS2, XIS3, and PIN
are shown as crosses colored black, red, green, blue, and cyan, respectively.
The spectral components, the BB, PL in the lower energy range, PL in the higher energy range,
SNR, and CXB + GRXE 
are shown in the histograms colored green, blue, red, cyan, and black, respectively.
Vertical error bars represent the $1\sigma$ level.
}\label{fig: wabs+snr+bb+pgpl+pgpl_allphase_07080410_corn+0p.ps}
\end{figure}

Among the many possible three-component models
we used only the ${\rm PL} + {\rm BB}$ and ${\rm BB} + {\rm BB}$ models
for the lower energy region,
because they are familiar models for AXPs below $\sim 10$ keV.
When we applied a ${\rm PL}$ model to the higher energy region,
the ${\rm PL} + {\rm BB} + {\rm PL}$ model was found to produce a good fit
(table \ref{tab: good fits} and
figure \ref{fig: wabs+snr+bb+pgpl+pgpl_allphase_07080410_corn+0p.ps}).
Although the ${\rm BB} + {\rm BB} + {\rm PL}$ model also produced a reasonable fit,
the radius of the BB component with the lower temperature became too large
in comparison with the radius of neutron stars ($\sim 10$ km)
(table \ref{tab: good fits}).

In the case of the ${\rm PL} + {\rm BB} + {\rm PL}$ model
the photon index for the hard X-ray component was
$1.62^{+0.05}_{-0.05}$(stat)$^{+0.16}_{-0.17}$(syst) (table \ref{tab: good fits}).
Since the photon index could mimic that of thermal bremsstrahlung
below the exponential cutoff energy
\footnote{
The slope of a thermal bremsstrahlung spectrum
below the exponential cutoff energy
exhibits two different dependencies on plasma temperature (kT) in the energy range of the PIN (10 - 50 keV).
These are the Gaunt factor and a round-off of
the exponential cutoff ($\exp(- E/kT)$).
Details of the former dependency were shown in \citet{Kellogg Baldwin Koch 1975}
and the result was included in the ``bremss'' model in XSPEC.
},
we tried applying a ${\rm PL} + {\rm BB} + {\rm TB}$ model.
It resulted in a comparatively good fit (table \ref{tab: good fits}).
The ${\rm BB} + {\rm BB} + {\rm TB}$ model was also good, although there was
a similar problem in the blackbody radius as in the case of
the ${\rm BB} + {\rm BB} + {\rm PL}$ model.
The results of the spectral fits are summarized in table \ref{tab: good fits}.

\begin{longtable}{lll}
  \caption{Fits to the phase-averaged total spectrum of 1E 1841$-$045}\label{tab: good fits}
\endfirsthead
\endhead
\endfoot
\endlastfoot

  \hline
         \multicolumn{3}{c}{BB + PL + PL} \\ \hline
parameter   & 
               units    &
             value $\pm$ stat$^\dagger$ $\pm$ syst$^\ddagger$  \\ \hline

$N_{\rm H}$ &
     10$^{22}$ cm$^{-2}$ &
$2.866^{+0.020}_{-0.020}$$^{+0.008}_{-0.010}$  \\

$kT_{\rm BB}$          & 
           keV             &
$0.536^{+0.015}_{-0.014}$$^{+0.027}_{-0.026}$  \\

$R_{BB}$      & 
           km @ 8.5 kpc       &
 $3.16^{+0.35}_{-0.34}$$^{+0.46}_{-0.45}$ \\

$\Gamma_{\rm low}$    &
                           &
$4.99^{+0.29}_{-0.29}$$^{+0.28}_{-0.30}$ \\

${\rm Flux}_{\rm low}$ $^*$ &
  $10^{-12}$ erg/cm$^2$/s @ 1 -- 50 keV  &
$46.5^{+2.4}_{-2.3}$$^{+1.1}_{-1.1}$ \\

$\Gamma_{\rm high}$    &
                           &
$1.62^{+0.05}_{-0.05}$$^{+0.16}_{-0.17}$ \\

${\rm Flux}_{\rm high}$ $^*$ &
  $10^{-12}$ erg/cm$^2$/s @ 1 -- 50 keV  &
$43.7^{+0.9}_{-0.9}$$^{+3.0}_{-2.0}$ \\

\multicolumn{3}{c}{$\chi^2$/d.o.f. = 2530 / 2154 = 1.175} \\

  \hline
         \multicolumn{3}{c}{BB + PL + TB} \\ \hline
$N_{\rm H}$ &
     10$^{22}$ cm$^{-2}$ &
$2.851^{+0.019}_{-0.018}$$^{+0.009}_{-0.008}$ \\

$kT_{\rm BB}$     & 
           keV             &
$0.569^{+0.014}_{-0.014}$$^{+0.015}_{-0.019}$ \\

$R_{\rm BB}$      & 
           km @ 8.5 kpc       &
$2.60^{+0.29}_{-0.28}$$^{+0.39}_{-0.28}$ \\

$\Gamma$    &
                           &
$4.56^{+0.26}_{-0.25}$$^{+0.24}_{-0.19}$ \\

Flux $^*$ &
  $10^{-12}$ erg/cm$^2$/s @ 1 -- 50 keV  &
$48.2^{+1.9}_{-1.8}$$^{+0.6}_{-0.6}$ \\

$kT_{\rm TB}$   & 
           keV             &
$51.7^{+14.1}_{-8.8}$$^{+68.0}_{-22.1}$ \\

${\rm EM}_{\rm TB}$($= n^2 V$)  & 
         $10^{57}$ cm$^{-3}$ @ 8.5 kpc     &
$7.47^{+0.19}_{-0.22}$$^{+0.73}_{-0.02}$ \\

\multicolumn{3}{c}{$\chi^2$/d.o.f. = 2534 / 2154 = 1.176} \\

  \hline
         \multicolumn{3}{c}{BB + BB + PL} \\ \hline

$N_{\rm H}$ &
     10$^{22}$ cm$^{-2}$ &
$2.811^{+0.013}_{-0.009}$$^{+0.007}_{-0.007}$ \\

$kT_{{\rm BB}_1}$          & 
           keV             &
$0.211^{+0.009}_{-0.009}$$^{+0.010}_{-0.010}$ \\

$R_{{\rm BB}_1}$      & 
           km @ 8.5 kpc       &
$55.7^{+10.7}_{-8.3}$$^{+9.4}_{-7.5}$ \\

$kT_{{\rm BB}_2}$          & 
           keV             &
$0.505^{+0.008}_{-0.011}$$^{+0.023}_{-0.022}$ \\

$R_{{\rm BB}_2}$      & 
           km @ 8.5 kpc       &
$4.72^{+0.30}_{-0.27}$$^{+0.47}_{-0.44}$ \\

$\Gamma$    &
                           &
$1.70^{+0.04}_{-0.03}$$^{+0.14}_{-0.14}$ \\

Flux $^*$  &
  $10^{-12}$ erg/cm$^2$/s @ 1 -- 50 keV  &
$44.3^{+0.8}_{-0.7}$$^{+2.9}_{-1.8}$ \\

\multicolumn{3}{c}{$\chi^2$/d.o.f. = 2541 / 2154  = 1.180 } \\

  \hline
         \multicolumn{3}{c}{BB + BB + TB} \\ \hline

$N_{\rm H}$ &
     10$^{22}$ cm$^{-2}$ &
$2.799^{+0.015}_{-0.014}$$^{+0.005}_{-0.004}$ \\

$kT_{{\rm BB}_1}$          & 
           keV             &
$0.228^{+0.009}_{-0.009}$$^{+0.007}_{-0.009}$ \\

$R_{{\rm BB}_1}$      & 
           km @ 8.5 kpc       &
$44.1^{+6.9}_{-5.2}$$^{+5.8}_{-3.6}$ \\

$kT_{{\rm BB}_2}$          & 
           keV             &
$0.543^{+0.011}_{-0.010}$$^{+0.016}_{-0.021}$ \\

$R_{{\rm BB}_2}$      & 
           km @ 8.5 kpc       &
$4.08^{+0.24}_{-0.22}$$^{+0.43}_{-0.29}$ \\

$kT_{\rm TB}$    &
              keV           &
$36.0^{+4.7}_{-4.0}$$^{+26.0}_{-12.6}$ \\

${\rm EM}_{\rm TB}$($= n^2 V$) &
         $10^{57}$ cm$^{-3}$ @ 8.5 kpc     &
$8.03^{+0.09}_{-0.09}$$^{+0.27}_{-0.01}$ \\

\multicolumn{3}{c}{$\chi^2$/d.o.f. =  2557 / 2154   = 1.187} \\ \hline 

\multicolumn{3}{l}{$^\dagger$ Statistical error of 1$\sigma$. } \\
\multicolumn{3}{l}{$^\ddagger$ Systematic error due to the 3\% uncertainty in
the non-X-ray background of the PIN\footnotemark[\ref{fn: suzaku-memo}].} \\
\multicolumn{3}{l}{$^*$ Unabsorbed flux.} \\

\end{longtable}

\subsection{Phase-resolved Spectroscopy}\label{subsec: Phase-resolved Spectroscopy}

\begin{figure}
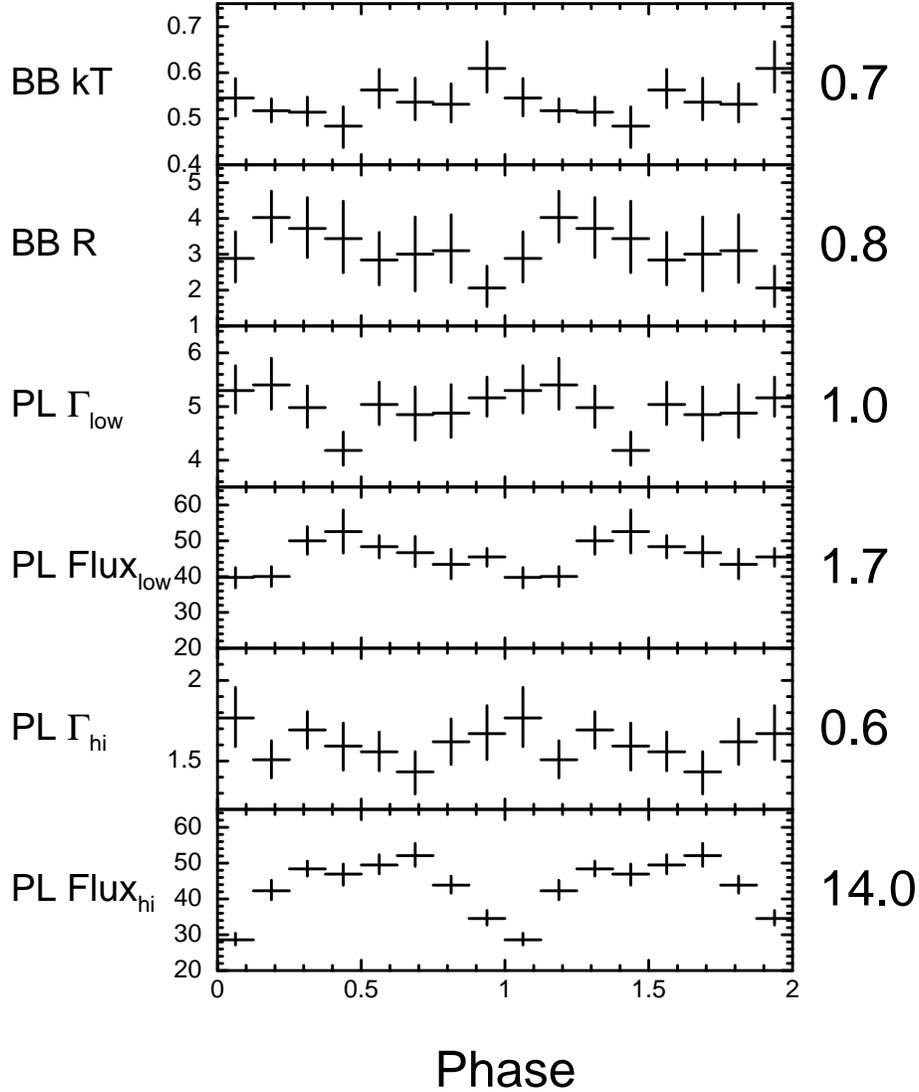

  \begin{center}

    \FigureFile(120mm,120mm){figure4.ps}
  \end{center}
  \caption{Variation of the spectral parameters along pulse phases,
	when the spectra were fitted with a BB + PL + PL model.
	The panels from the top to the bottom show
	the temperature of the BB component,
	the radius of the BB emission region on the neutron star surface,
	the photon index of the PL component in the lower energy range,
	the unabsorbed flux of the PL component in the lower energy range,
	the photon index of the PL component in the higher energy range
	and
	the unabsorbed flux of the PL component in the higher energy range.
	The fluxes are in an energy range of 1.0 -- 50.0 keV.
	The temperatures, radii and fluxes are shown in units of keV, km
	and $\times 10^{-12}$ erg s$^{-1}$ cm$^{-2}$, respectively.
	Vertical error bars represent a $1\sigma$ level.
	At the right of each panel the reduced $\chi^2$ is shown
	for the case where the profile was fitted with a constant model and
	the degree of freedom was 7.
	}\label{fig: specvar_phase_1sigma_07080410_071123-155142-25742.ps}
\end{figure}

\begin{figure}
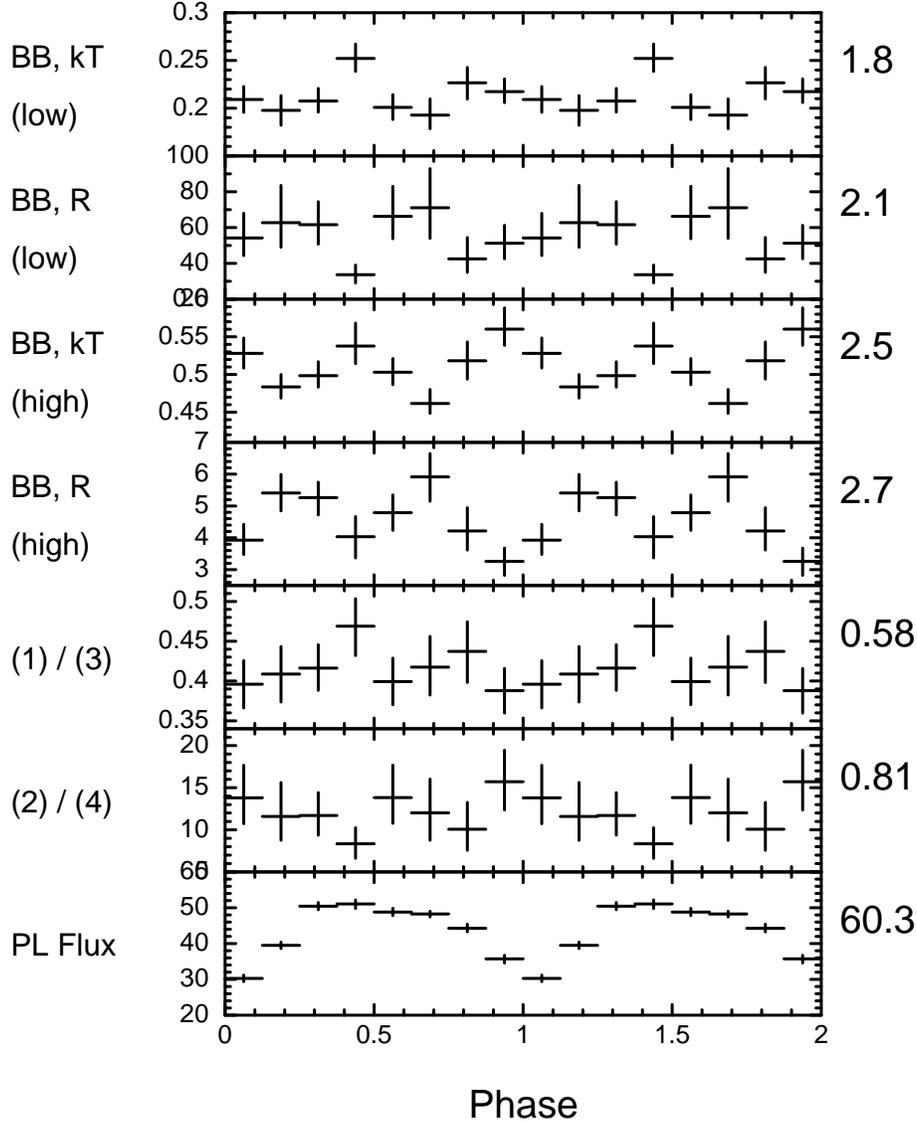

  \begin{center}

    \FigureFile(120mm,120mm){figure5.ps}
  \end{center}
  \caption{Variation of spectral parameters along pulse phases,
	when the spectra were fitted with a BB + BB + PL model.
	Here, the photon indices of the power-law components were fixed with the
	best value obtained from the phase-averaged spectroscopy
	 (table \ref{tab: good fits}).
	The top two panels show the temperature
  	and radius of the emission region on the neutron star surface of the BB component
	in the lower energy range.
	The third and fourth panels show those in the higher energy range.
	The fifth and sixth panels show the ratios of
	the first to the third panel and the second to the fourth, respectively.
	The bottom panel shows the unabsorbed flux of the PL component
	in an energy range of 1.0 -- 50.0 keV.
	The temperatures, radii and fluxes are shown in units of keV, km
	and $\times 10^{-12}$ erg s$^{-1}$ cm$^{-2}$, respectively.
	Vertical error bars represent a $1\sigma$ level.
	At the right of each panel the reduced $\chi^2$ is shown
	for the case where the profile was fitted with a constant model and
	the degree of freedom was 7.

	}
\label{fig: specvar_phase_1sigma_wabs+snr+bb+bb+pgpl_ph-ga-hi-fx_08bin_07080435_080128-162352-16274_bbbol_for_paper2.ps}
\end{figure}

We divided the event data into 8 equally wide phase intervals
and constructed the phase-resolved spectra
by the same procedure shown in subsection \ref{subsection: Preparation}.
We rebinned the PI spectra of the XISs so that
$N_{\rm min}$ (see subsection \ref{subsection: Preparation})
was 50, 100 and 50 for the energies below 1.1, between 1.1 and 4.0 and
above 4.0 keV, respectively.
We used $N_{\rm min} = 300$ counts for the PIN spectra.
We fitted each phase-resolved spectrum by fixing
the constant parameters during all phases to the best values obtained
by the phase-averaged spectroscopy 
(sections \ref{subsection: Preparation} and \ref{subsection: Phase-averaged Spectroscopy}).
The fixed parameters were the gains, the normalizations of the XISs and PIN,
the normalization of the SNR component and the column density
for the line of sight ($N_H$).
Figures \ref{fig: specvar_phase_1sigma_07080410_071123-155142-25742.ps}
and 
\ref{fig: specvar_phase_1sigma_wabs+snr+bb+bb+pgpl_ph-ga-hi-fx_08bin_07080435_080128-162352-16274_bbbol_for_paper2.ps}
show the variations of the parameters when the spectra were
fitted with the ${\rm BB} + {\rm PL} + {\rm PL}$
and ${\rm BB} + {\rm BB} + {\rm PL}$ models, respectively.
For the latter model
the photon indices of the power-law components
were constant during all phases within a 90\% confidence level (C.L.).
Then, we fixed those indices with the best value
obtained by the phase-averaged spectroscopy
(table \ref{tab: good fits}).
In these figures the vertical bars represent a statistical error
of $1\sigma$, not including the errors caused by the
systematic uncertainty of the non-X-ray background.

\subsection{Spectroscopy of the Pulsed Component}\label{subsection: spec pulse tot}

\begin{figure}
  \begin{center}

    \FigureFile(150mm,150mm){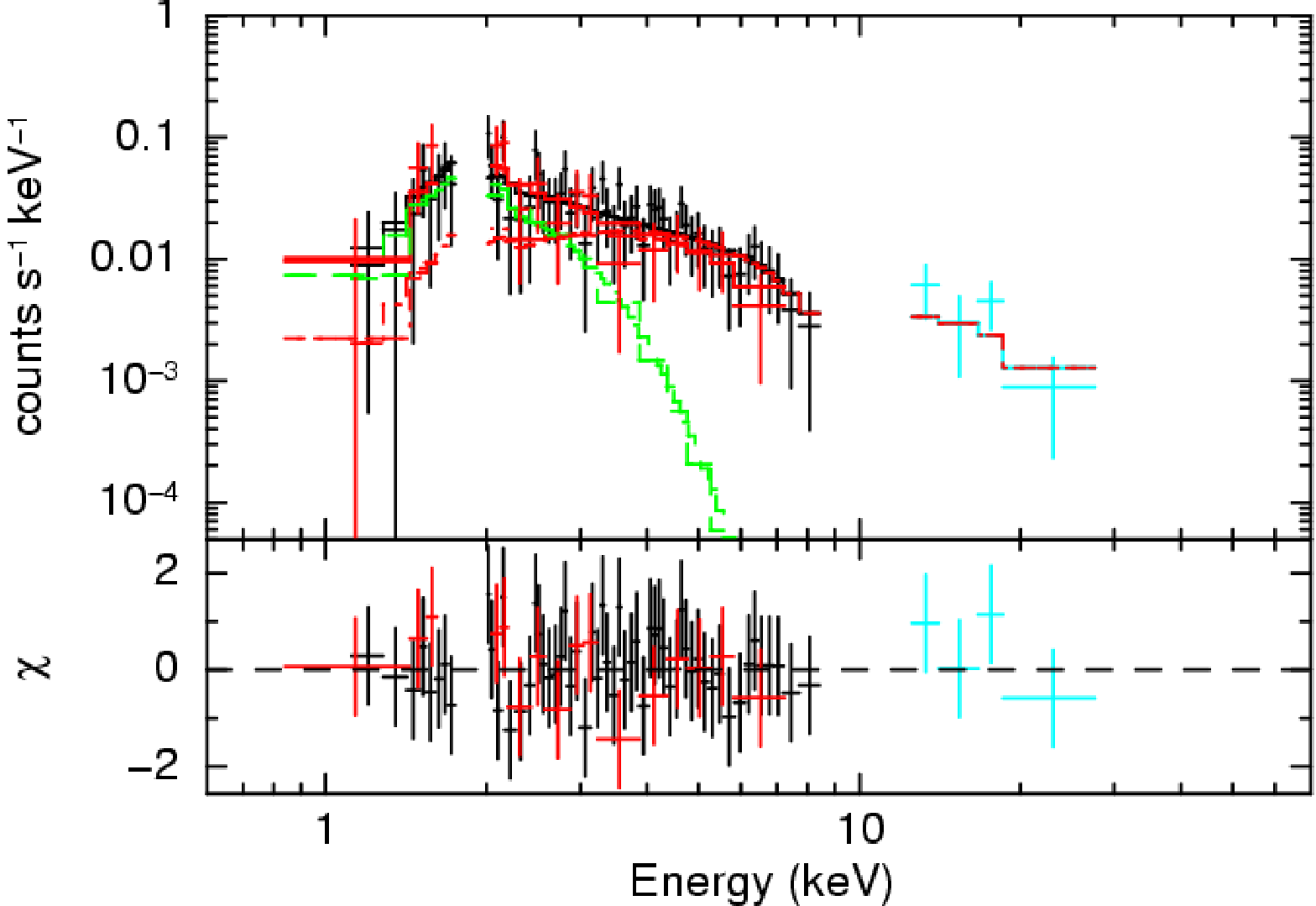}

  \end{center}
  \caption{Spectrum of the pulsed component fitted with a PL + BB model.
The data points from the XIS-FI, XIS-BI and PIN
are shown as crosses colored black, red and cyan, respectively.
The BB and PL spectral components are shown as green and red histograms,
respectively. Vertical error bars represent a $1\sigma$ level.
}\label{fig: wabs+bb+pgpl_bin_pls_08092202_corn+0p_paper_rot.ps}
\end{figure}

The spectrum of the pulsed component was produced
by subtracting the off-pulse spectrum in the phase range 0.0 -- 0.1
from the spectrum with a range between 0.1 and 1 using \texttt{mathpha}.
We constructed a XIS-FI spectrum by adding the spectra of XIS0, XIS2 and XIS3,
and also a response by using \texttt{marfrmf} and \texttt{addrmf}.
We rebinned the PI spectra of XIS-FI, XIS-BI and PIN so that
$N_{\rm min}$ (see subsection \ref{subsection: Preparation})
was 500, 500 and 400, respectively.
We fitted the spectrum of the pulsed component
with ${\rm PL}$ and ${\rm BB} + {\rm PL}$ models
by fixing the $N_H$ to the best value
obtained from fitting the phase-averaged spectrum 
using the ${\rm BB} + {\rm PL} + {\rm PL}$ model
(subsection \ref{subsection: Phase-averaged Spectroscopy} and
table \ref{tab: good fits}).
Here, we averaged the gain parameters of XIS0, XIS2 and XIS3 to
obtain that of XIS-FI, and also renormalized and fixed
the parameters $c_{\rm XIS-BI}$ and $c_{\rm PIN}$.
The fitting results are shown in table \ref{tab: pulsed component fits}.
When the spectrum was fitted with the ${\rm PL}$ model,
there appeared to be a global concave residual,
although the statistics of the fit was still acceptable.
When it was fitted with the ${\rm PL} + {\rm BB}$ model,
this residual disappeared and the $\chi^2$ value
improved significantly with a F-test statistics of 9.80
(a probability of $1.7 \times 10^{-4}$).
Figure \ref{fig: wabs+bb+pgpl_bin_pls_08092202_corn+0p_paper_rot.ps} shows
the spectrum of the pulsed component fitted with the ${\rm PL} + {\rm BB}$ model.

\begin{longtable}{lll}
  \caption{Fits to the spectrum of the pulsed component
of 1E 1841$-$045}\label{tab: pulsed component fits}
\endfirsthead
\endhead
\endfoot
\endlastfoot

  \hline
         \multicolumn{3}{c}{PL $^\dagger$} \\ \hline

parameter   & 
               units    &
             value $\pm$ stat($1\sigma$)    \\ \hline
$\Gamma$ &
                           &
     $2.45^{+0.20}_{-0.21}$ \\

${\rm Flux}$ $^*$ &
  $10^{-12}$ erg/cm$^2$/s @ 1 -- 50 keV  &
     $13.7^{+1.1}_{-0.9}$ \\
  
\multicolumn{3}{c}{$\chi^2$/d.o.f. = 47.1 /75 = 0.627} \\

  \hline
         \multicolumn{3}{c}{BB + PL $^\dagger$} \\ \hline

$kT_{\rm BB}$          & 
           keV             &
$0.37^{+0.06}_{-0.05}$ \\
  
$R_{BB}$      & 
           km @ 8.5 kpc       &
$5.3^{+2.7}_{-1.7}$ \\

$\Gamma$ &
                           &
     $1.35^{+0.30}_{-0.25}$ \\

${\rm Flux}$ $^*$ &
  $10^{-12}$ erg/cm$^2$/s @ 1 -- 50 keV  &
     $21.8^{+6.2}_{-5.1}$ \\

\multicolumn{3}{c}{$\chi^2$/d.o.f. = 37.1 / 73 = 0.508} \\ \hline
\multicolumn{3}{l}{$^\dagger$ $N_{\rm H}$ is fixed to the best value obtained
by the phase-averaged spectral analysis} \\
\multicolumn{3}{l}{of the total emission using the ${\rm BB} + {\rm PL} + {\rm PL}$ model
 (see subsection \ref{subsection: Phase-averaged Spectroscopy} and table \ref{tab: good fits}).} \\
\multicolumn{3}{l}{$^*$ Unabsorbed flux.} \\
\end{longtable}

\subsection{Phase-resolved Spectroscopy of the Pulsed Component}
\label{subsection: phase-res spec of pulse comp}

\begin{figure}
  \begin{center}
    \FigureFile(150mm,150mm){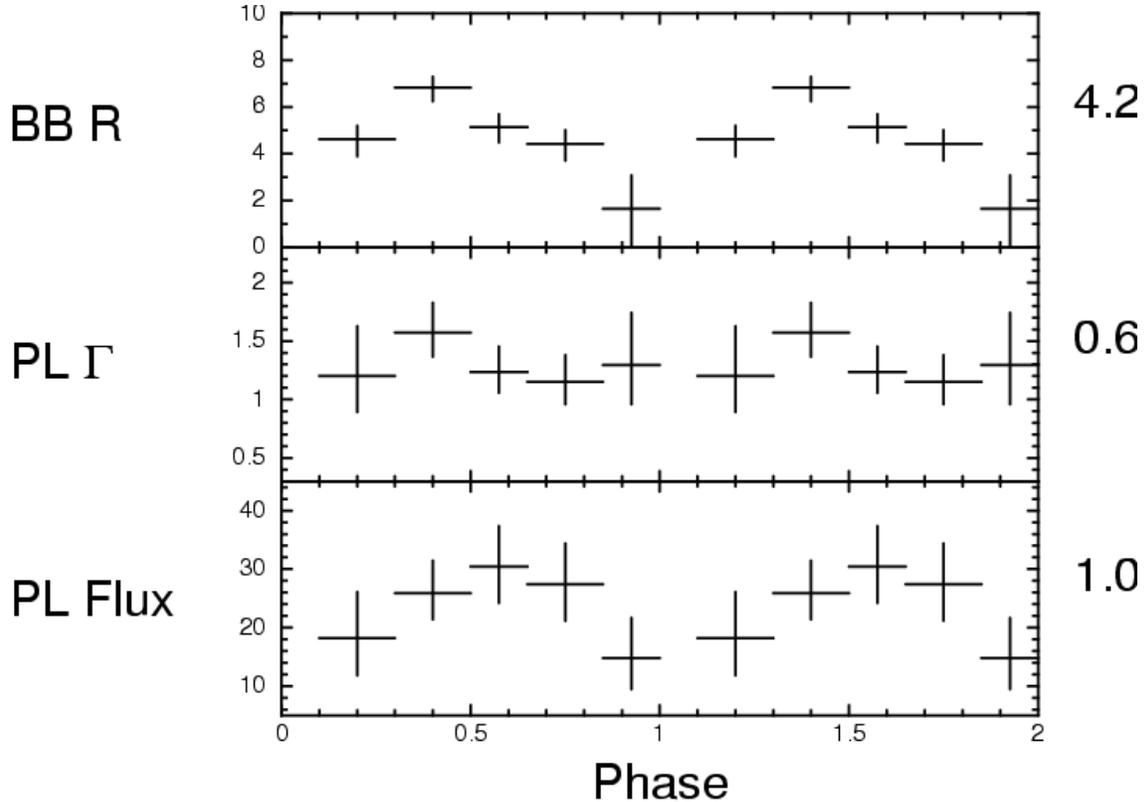}

  \end{center}
  \caption{Variation of spectral parameters along pulse phase,
when the pulsed spectra were fitted with a BB + PL model with the temperature fixed to
the best fitting value (see text).
The panels from the top to the bottom show the radius of the BB emission region on
the neutron star surface, the photon index of the PL component
and the unabsorbed flux of the PL component.
The fluxes are in an energy range of 1.0 -- 50.0 keV. The radii and fluxes are shown 
in units of km and $\times 10^{-12}$ erg s$^{-1}$ cm$^{-2}$, respectively.
Vertical error bars represent a $1\sigma$ level.
Horizontal bars show the phase binning described in section 
\ref{subsection: phase-res spec of pulse comp}.
At the right of each panel the reduced $\chi^2$ value is shown
for the case where the profile was fitted with
a constant model and the degree of freedom was 4.
}\label{fig: wabs+bb+pgpl_ktfx_specvar_phase_1sigma_08092202_080926-110336-18926_paper_rot.ps}
\end{figure}

We divided the data into the following phase intervals:
off-pulse (0.0 -- 0.1),
rising wing of the first pulse (0.1 -- 0.3),
top of the first pulse (0.3 -- 0.5),
valley (0.5 -- 0.65),
top of the second pulse (0.65 -- 0.85)
and trailing wing of the second pulse (0.85 -- 1.0).
Phase-resolved spectra of the pulsed component were
constructed by subtracting the off-pulse spectrum
for intervals other than the off-pulse interval.
We produced XIS-FI spectra by the procedure
described in subsection \ref{subsection: spec pulse tot}.
We rebinned the PI spectra of XIS-FI, XIS-BI and PIN so that
$N_{\rm min}$ (see subsection \ref{subsection: Preparation})
was 100, 100 and 50, respectively.
We first fitted all the phases with ${\rm BB} + {\rm PL}$ models
by fixing the $N_H$ to the best value
obtained by fitting the phase-averaged spectrum using 
the ${\rm BB} + {\rm PL} + {\rm PL}$ model
(subsection \ref{subsection: Phase-averaged Spectroscopy} and
table \ref{tab: good fits})
and allowing the other parameters to vary.
In this case fits for all phases were acceptable within the
90\% C.L. and the parameters showed no significant deviation
from the mean values.
We then fixed the blackbody temperature to the best value
obtained by fitting the pulsed component spectrum using the
${\rm BB} + {\rm PL}$ model (subsection \ref{subsection: spec pulse tot}
and table \ref{tab: pulsed component fits}).
In this case all of the fits were also acceptable within the 90\% C.L.
The variations of the parameters are shown in figure
\ref{fig: wabs+bb+pgpl_ktfx_specvar_phase_1sigma_08092202_080926-110336-18926_paper_rot.ps}.

\subsection{Comparison of the Suzaku spectral results with other high-energy instruments}
\label{suzaku-integral}

\begin{figure}
  \begin{center}
    \FigureFile(150mm,150mm){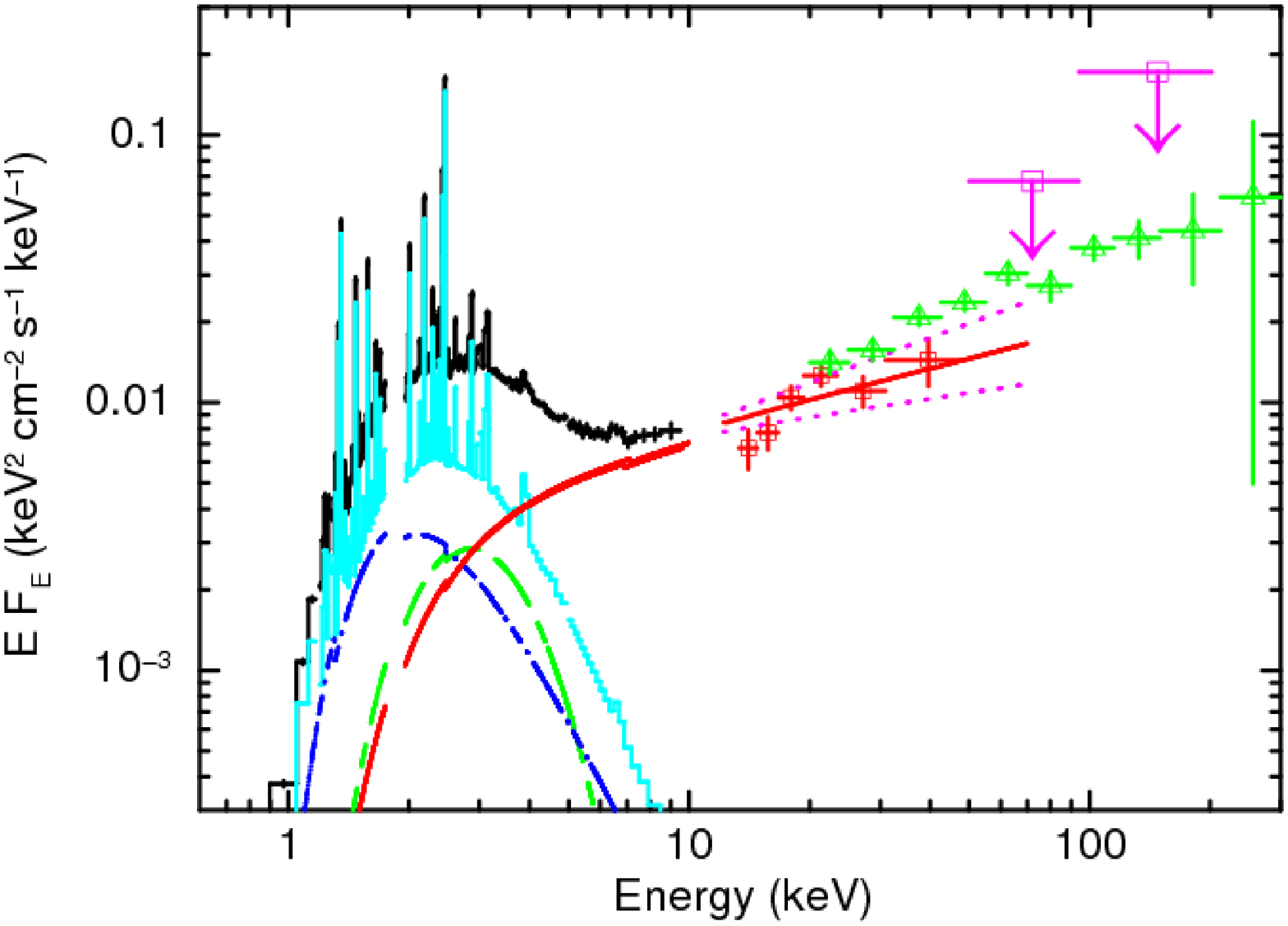}
  \end{center}
  \caption{A $\nu F_\nu$ plot for phase-averaged total (pulsed + DC)
spectrum with interstellar absorption obtained by the Suzaku observation
compared with that of INTEGRAL IBIS ISGRI \citep{Kuiper den Hartog Hermsen 2008}.
The vertical and horizontal axes are energy-scaled flux $E F_E$ and energy in
units of keV$^2$ cm$^{-2}$ s$^{-1}$ keV$^{-1}$ and keV, respectively.
The phase-averaged spectrum of Suzaku is produced by unfolding in the BB + PL + PL
with the best fitting parameters of table \ref{tab: good fits}.
The phase-averaged fluxes of XIS-FI and PIN are shown as 
black crosses and red crosses with square symbols, respectively.
The upper limits of GSO are shown as magenta downward arrows
corresponding to the 2\% uncertainty of the non-X-ray background.
The blackbody, power-laws at lower and higher energy and
supernova remnant components of Suzaku are shown as
green dashed, blue dot-dashed, red solid and cyan solid lines,
respectively.
The range of the systematic uncertainty in the PIN for the high energy power-law component
is shown as two magenta dotted lines.
All vertical error bars represent a $1\sigma$ level.
For comparison the INTEGRAL ISGRI 20$-$300 keV
spectrum \citep{Kuiper den Hartog Hermsen 2008} is also plotted
as green crosses with triangle symbols.
}\label{fig: nufnu_total}
\end{figure}

\begin{figure}
  \begin{center}
    \FigureFile(150mm,150mm){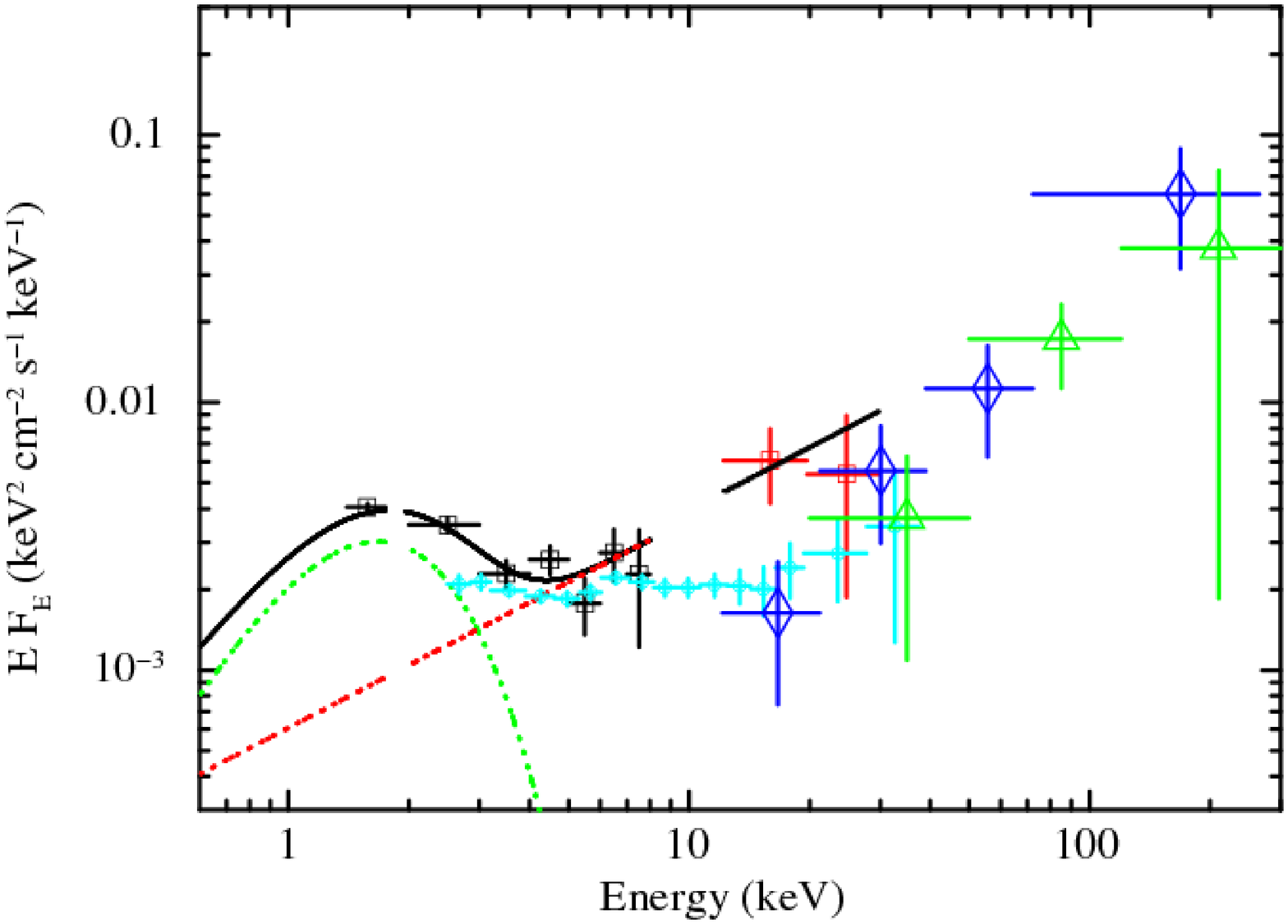}

  \end{center}
  \caption{An unabsorbed $\nu F_\nu$ plot for the pulsed component
obtained from the Suzaku observation
compared with those of RXTE PCA, RXTE HEXTE and INTEGRAL IBIS ISGRI 
\citep{Kuiper et al. 2006, Kuiper den Hartog Hermsen 2008}.
The vertical and horizontal axes are energy-scaled flux $E F_E$ and energy in
units of keV$^2$ cm$^{-2}$ s$^{-1}$ keV$^{-1}$ and keV, respectively.
The pulsed component spectrum of Suzaku is produced by unfolding in the BB + PL models
with the best fitting parameters of table \ref{tab: pulsed component fits}.
The total model, blackbody and power-law components are shown as
black solid, green and red dotted lines, respectively.
The XIS-FI and PIN data are shown as black and red crosses with square symbols.
All vertical error bars represent a $1\sigma$ level.
For comparison 
RXTE PCA, RXTE HEXTE and INTEGRAL IBIS ISGRI 
\citep{Kuiper et al. 2006, Kuiper den Hartog Hermsen 2008}
are shown as crosses with cyan small circle,
blue diamond and green triangle symbols, respectively.
}\label{fig: nufnu_pls}
\end{figure}

Finally, we made $\nu F_\nu$ plots
for the total (phase-averaged spectrum) and pulsed components
(figures \ref{fig: nufnu_total} and \ref{fig: nufnu_pls}, respectively).
Here, the phase-averaged and pulsed component Suzaku spectra are
unfolded by the BB + PL + PL and the BB + PL models
with the best fitting parameters of tables \ref{tab: good fits}
and \ref{tab: pulsed component fits}, respectively.
In figure \ref{fig: nufnu_total}
the flux of around $20 - 50$ keV for the Suzaku data
is marginally consistent with that of INTEGRAL IBIS ISGRI.
The photon index of the PL component in the higher energy range
($1.62^{+0.05}_{-0.05}$(stat)$^{+0.16}_{-0.17}$(syst))
is also marginally consistent with that of INTEGRAL ($1.32 \pm 0.11$),
where the difference corresponds to $1.5 \,\, \sigma$
if both statistical and systematic errors are assumed to be $1\sigma$
and combined as a root-mean-squared value.
In figure \ref{fig: nufnu_pls} the pulsed component flux of the XISs
is consistent with that of RXTE PCA in the 2.5$-$8 keV range.
However, the flux bin of the PIN in the 12$-$20 keV range is
marginally larger than that of RXTE PCA
by a factor of 2.8 and with a significance of 1.9$\sigma$ level. 

\section{Discussion}

In the previous Chandra observations \citep{Morii et al. 2003}
the spectrum of the AXP 1E 1841$-$045 was
described well in the energy range
$0.6$ -- $7.0$ keV by a combination of a blackbody (BB) and
a power-law (PL) function or that of two blackbodies.
In the BB + PL model the temperature and photon index were
determined to be $kT_{\rm BB} = 0.44 \pm 0.02$ keV and $\Gamma = 2.0 \pm 0.3$.
Since the photon index was the flattest among AXPs and close to
those of SGRs ($\sim 2$), \citet{Morii et al. 2003} concluded
that this AXP was the closest object to SGRs among AXPs.
Nonetheless, owing to the hard X-ray component above $\sim 20$ keV 
discovered by \citet{Kuiper Hermsen Mendez 2004}
these conclusions had to be modified, because
a substantial amount of the hard X-ray component is thought to
be included in the Chandra spectrum.
This expectation has been confirmed by our Suzaku observation and
an additional component in the hard X-ray band turns out to be necessary.

When the spectrum was modeled with a BB + PL + PL model, it
produced the best fit among the models we applied
(table \ref{tab: good fits}).
This modeling was the most plausible because the hard X-ray spectrum was well
modeled by a power-law function above $\sim 20$ keV in the result of
\citet{Kuiper Hermsen Mendez 2004}.
On the other hand, we found that the photon index of the lower energy range
$\Gamma_{\rm low} = 4.99^{+0.29}_{-0.29}$(stat)$^{+0.28}_{-0.30}$(syst)
is quite different from that of the previous Chandra result
(section \ref{sec: Introduction} and at the beginning of this section).
This discrepancy was caused by the different energy coverage of the spectroscopic data
between Suzaku and other satellites.
While the spectra obtained by the other satellites could be fitted by only one or two continuum components,
we could fit this AXP spectrum satisfactorily only by three continuum component
due to the much wider energy coverage of our Suzaku measurements.
Therefore, Suzaku observations of AXPs and SGRs
are important for determination of spectra of magnetars.
In particular, the relationship between spectral properties and spin-down rates
among AXPs and SGRs obtained by \citet{Marsden White 2001} must be reexamined by Suzaku observations.
Ironically, as shown in figure \ref{fig: wabs+snr+bb+pgpl+pgpl_allphase_07080410_corn+0p.ps}
the PL component with the steep index does not contribute to the hard tail shape
but rather the excess component in the lower energy range.
Therefore, the physical meaning of this PL component should be reconsidered.

The photon index of the hard X-ray component in the BB + PL + PL model
was similar to that of thermal bremsstrahlung below the exponential cutoff energy
(see subsection \ref{subsection: Phase-averaged Spectroscopy}).
Thus, a ${\rm PL} + {\rm BB} + {\rm TB}$ model could also fit the spectrum
(table \ref{tab: good fits}).
The temperature of the thermal bremsstrahlung was 
$kT_{\rm TB} = 51.7^{+14.1}_{-8.8}$(stat)$^{+68.0}_{-22.1}$(syst) keV,
implying the existence of a high energy cut-off above the energy range of the PIN.
Nonetheless 
the INTEGRAL ISGRI data, collected over many years (figure \ref{fig: nufnu_total})
contradict this model,
because the data shows that the cut-off energy is above about $200 - 300$ keV.
However, there still remain a possibility that
the temperature of the thermal bremsstrahlung was low during the Suzaku observation
due to short time scale variability.

In the pulsed component spectrum of figure \ref{fig: nufnu_pls}
the flux of the XISs is consistent with that of RXTE PCA.
Nonetheless, that of the PIN for $12-20$ keV is marginally higher than
that of RXTE PCA and HEXTE. This discrepancy could be another indication for
the spectral variability of the hard tail component over a short time scale.
Further Suzaku observation of this source will be needed to confirm this kind of variability in the hard tail.

We showed for the first time analysis of phase-resolved spectra for
the total (including the DC component) and the pulsed component (excluding the DC component)
in subsection \ref{subsection: spec pulse tot} and
\ref{subsection: phase-res spec of pulse comp}, respectively.
The shift of the peak phases for the blackbody radii and the PL fluxes
in the pulsed component spectra
(figure \ref{fig: wabs+bb+pgpl_ktfx_specvar_phase_1sigma_08092202_080926-110336-18926_paper_rot.ps})
is consistent with the peak shifts of pulse profiles in figure \ref{fig: prof+hr.ps}.
This shift indicates the existence of
two emission regions: one is a hot spot on the neutron star surface
with blackbody emission and the other is of non-thermal origin
rather than two hot spots on the surface.
The latter possibility was first put forward by \citet{Morii et al. 2003}
and turns out to be unlikely on the basis of this study.
This interpretation is also confirmed
by the peak shift of the two PL fluxes 
in figure \ref{fig: specvar_phase_1sigma_07080410_071123-155142-25742.ps}
(the forth and the bottom panels).
Here, the PL component in the lower energy range
substantively represents the soft excess component as discussed above.

The photon index of the pulsed component (table \ref{tab: pulsed component fits})
is similar to
those of several young and middle-aged rotation powered pulsars
such as Vela pulsar, PSR B1509$-$58 and PSR J1846$-$0258
(see \cite{Kuiper et al. 2006}, \cite{Kuiper Hermsen 2009}
and references therein).
Then, the outer-gap or slot-gap models describing the production of gamma-ray emission
of rotation-powered pulsars are promising to explain
the hard X-ray emission of magnetars.

\begin{figure}
  \begin{center}
    \FigureFile(150mm,150mm){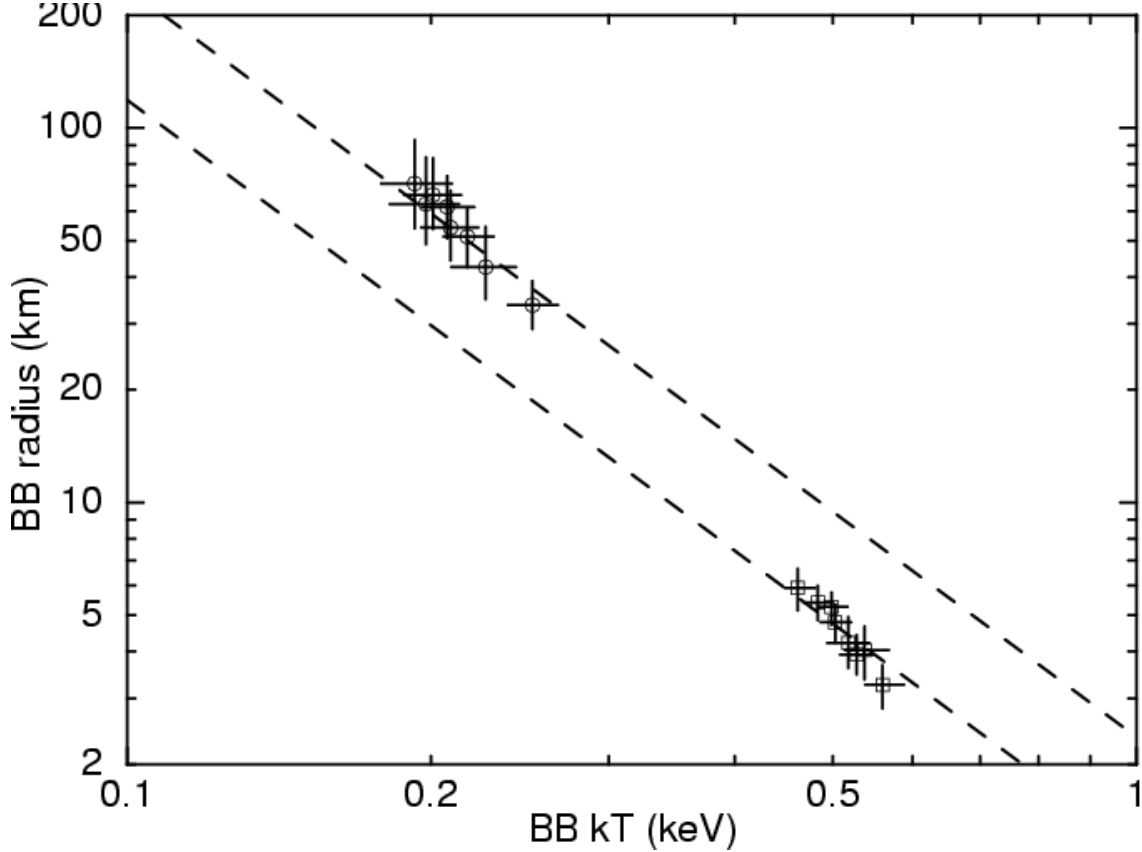}
  \end{center}
  \caption{Anti-correlation between radii ($R$) and temperatures ($kT$) for
the blackbody components with lower and higher temperatures,
when the phase-resolved spectra were fitted with a BB + BB + PL model.
The crosses with circular symbols and those with square symbols denote lower and higher temperatures, respectively.
Error bars represent $1\sigma$ levels.
The dashed lines represent a constant blackbody bolometric luminosity, that is
$R \propto (kT)^{-2}$.
}
\label{fig: kt-km_wabs+snr+bb+bb+pgpl_ph-ga-hi-fx_08bin_07080435.ps}
\end{figure}

When the phase-averaged total spectrum was fitted with the BB + BB + PL model,
the area of the emission region of the blackbody component
with the lower temperature
was too large in comparison with the surface area of neutron stars (table \ref{tab: good fits}).
Therefore, these models are physically unacceptable at first glance.
Nonetheless, \citet{Nakagawa et al. 2009} showed
interesting relationships among the parameters of the BB + BB model
in the spectra of magnetars below about 10 keV,
which may offer clues to the underlying physics of magnetars.
For example, \citet{Ozel 2001} showed that the shape
of the spectra of magnetars is deformed by cyclotron resonant scattering
in the magnetosphere and the resulting spectra are
well fitted by BB + BB and BB + PL models.
We then checked whether the phase-resolved spectra of 1E 1841$-$045
also follow this relationship.
In what follows, we will treat the BB + BB model simply
as an empirical model to represent the spectrum approximately.

When the spectra of each pulse phase interval was fitted with the BB + BB + PL model,
we found a significant anti-correlation
between the temperatures and radii of the two blackbody components
(figure \ref{fig: specvar_phase_1sigma_wabs+snr+bb+bb+pgpl_ph-ga-hi-fx_08bin_07080435_080128-162352-16274_bbbol_for_paper2.ps}).
This anti-correlation is clearly seen
in figure \ref{fig: kt-km_wabs+snr+bb+bb+pgpl_ph-ga-hi-fx_08bin_07080435.ps}.
It is similar to the anti-correlation reported by \citet{Nakagawa et al. 2009}
among phase-averaged spectra of various AXPs and SGRs.
\citet{Nakagawa et al. 2009} showed
that the blackbody radii ($R$) and temperatures ($kT$) follow
$R \propto (kT)^{-2}$, meaning that the bolometric luminosities of the blackbody
components are constant.
As indicated by the two dashed lines in figure
\ref{fig: kt-km_wabs+snr+bb+bb+pgpl_ph-ga-hi-fx_08bin_07080435.ps},
the $R$ and $kT$ of the two blackbody components with
the lower and higher temperature can be
fitted to a $R \propto (kT)^{-2}$ function at a 90\% C.L.
This is the first time such a relationship has been found
in the phase-resolved spectra of AXPs and SGRs.

\begin{figure}
  \begin{center}
    \FigureFile(150mm,150mm){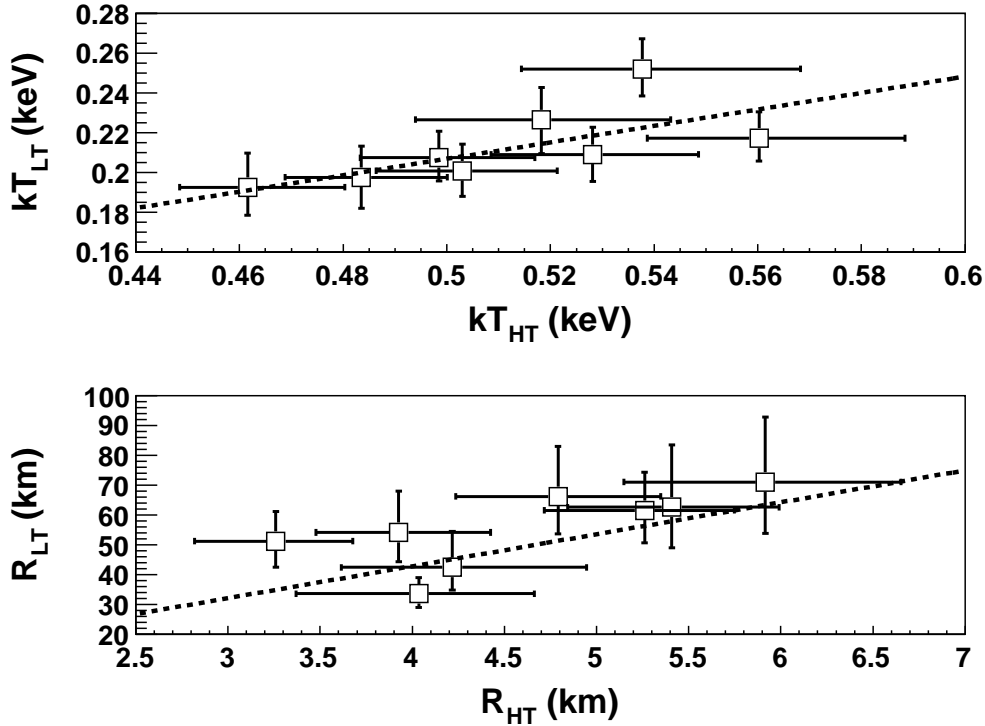}
  \end{center}
  \caption{Correlation between temperatures ($kT$) and radii ($R$) for
blackbody components with lower (LT) and higher (HT) temperatures,
when the phase-resolved spectra were fitted with a BB + BB + PL model.
Error bars represent $1\sigma$ levels.
The dotted lines shows $kT_{\rm LT} = 0.41 kT_{\rm HT}$
and $R_{\rm LT} = 10.7 R_{\rm HT}$.
}
\label{fig: rtfit.eps}
\end{figure}

We also tested other correlations between temperatures ($kT$) and radii ($R$) of
the two blackbodies reported by \citet{Nakagawa et al. 2009}.
These are  $kT_{\rm LT} / kT_{\rm HT} \sim 0.4$ and $R_{\rm LT} / R_{\rm HT} \sim 10$,
where the subscripts LT and HT denote
the lower and higher temperatures, respectively.
We found that these correlations also hold for the phase-resolved spectra of the AXP
(the fifth and sixth panels of
figure \ref{fig: specvar_phase_1sigma_wabs+snr+bb+bb+pgpl_ph-ga-hi-fx_08bin_07080435_080128-162352-16274_bbbol_for_paper2.ps} and figure \ref{fig: rtfit.eps})
for the first time in phase-resolved spectra of AXPs and SGRs.
In figure \ref{fig: rtfit.eps} the ratios are calculated to
be $kT_{\rm LT} / kT_{\rm HT} = 0.41 \pm 0.01$ and
$R_{\rm LT} / R_{\rm HT} = 10.7 \pm 0.8$ by fitting the points with 
a proportional expression.
In addition, we tested these correlations by calculating the Spearman rank-order correlation
coefficients ($r_s$; \cite{Press et al.(1992)}) for eight points with asymmetrical errors.
Here, the errors are taken into account by a Monte Carlo method with
a two dimensional Gaussian probability function with asymmetrical sigmas.
The results were $0.48^{+0.38}_{-0.51}$
and $0.43^{+0.41}_{-0.55}$ (90\% C.L.) respectively,
suggesting these values are moderately correlated.
These correlations suggest that
the spectral shape exhibited by blackbodies with two different temperatures
with such constraints can approximate the intrinsic spectral shape of magnetars.
This spectral shape is actually a self-similar function with two free parameters.

In summary, our Suzaku observation and analysis of
the AXP 1E 1841$-$045 showed that a BB + PL + PL model 
can best represent the phase-averaged spectrum over a wide energy range, simultaneously. 
Detailed phase-resolved spectroscopy was also carried out and
the existence of two emission regions was indicated, a hot spot with blackbody emission and another region with non-thermal emission.
The BB + BB + PL model was also statistically acceptable, although physically unacceptable.
Nonetheless, we found correlations between the radii and temperatures
of the two blackbodies in the phase-resolved spectra for the first time
in phase-resolved spectra of AXPs and SGRs.
These correlations may provide important insights into the underlying emission mechanisms of magnetars. 

\bigskip

We express our gratitude to L.~Kuiper for suggesting detailed revisions to this manuscript.
We are grateful to all the members of the Suzaku Science Working Group
and Suzaku Help. 
We appreciate Dai Takei at Rikkyo University for the Suzaku analysis.
Mikio Morii acknowledges support by a Grant-in-Aid for Young Scientists (B)(21740140)
and the Global COE Program, ``Nanoscience and
Quantum Physics Project of the Tokyo Institute of Technology.''


\begin{thebibliography}{}

\bibitem[Beloborodov \& Thompson(2007)]{Beloborodov Thompson 2007}
Beloborodov,~A.~M. \& Thompson,~C. 2007, ApJ, 657, 967

\bibitem[Borkowski et al.(2001)]{Borkowski et al. 2001}
Borkowski,~K.~J., Lyerly,~W.~J. \& Reynolds,~S.~P. 2001, ApJ, 548, 820

\bibitem[Buccheri et al. (1983)]{Buccheri+1983}
Buccheri,~R. {\it et al.} 1983, A\&A, 128, 245

\bibitem[Camilo et al.(2006)]{Camilo et al. 2006}
Camilo,~F. Ransom,~S.~M., Halpern,~J.~P., Reynolds,~J.,
Helfand,~D.~J., Zimmerman,~N. \& Sarkissian,~J. 2006, Nature, 442, 892

\bibitem[Camilo et al.(2007)]{Camilo et al. 2007}
Camilo,~F., Ransom,~S.~M., Halpern,~J.~P. \& Reynolds,~J. 2007, ApJ, 666, L93

\bibitem[den Hartog et al.(2008)]{den Hartog et al. 2008}
den Hartog,~P.~R., Kuiper,~L., Hermsen,~W., Kaspi,~V.~M.,
Dib,~R., Knodlseder,~J. \& Gavriil,~F.~P. 2008, A\&A, 489, 245

\bibitem[den Hartog, Kuiper \& Hermsen(2008)]{den Hartog Kuiper Hermsen 2008}
den Hartog,~P.~R., Kuiper,~L. \& Hermsen,~W. 2008, A\&A, 489, 263

\bibitem[Dib, Kaspi \& Gavriil(2008)]{Dib Kaspi Gavriil 2008}
Dib,~R., Kaspi,~V.~M. \& Gavriil,~F.~P. 2008, ApJ, 673, 1044

\bibitem[Hellier(1994)]{Hellier 1994}
Hellier,~C. 1994, MNRAS, 271, L21

\bibitem[Heyl \& Hernquist(2005)]{Heyl Hernquist 2005}
Heyl,~J.~S. \& Hernquist,~L. 2005, MNRAS, 362, 777

\bibitem[Kaneda et al.(1997)]{Kaneda et al. 1997}
Kaneda,~H., Makishima,~K., Yamauchi,~S., Koyama,~K., Matsuzaki,~K.
\& Yamasaki,~N.~Y. 1997, ApJ, 491, 638

\bibitem[Kellogg, Baldwin \& Koch(1975)]{Kellogg Baldwin Koch 1975}
Kellogg,~E., Baldwin,~J.~R. \& Koch,~D. 1975, ApJ, 199, 299

\bibitem[Kokubun et al.(2007)]{Kokubun et al. 2007}
Kokubun,~M. et al. 2007, PASJ, 59, S53 

\bibitem[Koyama et al.(2007)]{Koyama et al. 2007}
Koyama,~K. et al. 2007, PASJ, 59, S23 

\bibitem[Kriss et al.(1985)]{Kriss et al. 1985}
Kriss,~G.~A., Becker,~R.~H., Helfand,~D.~J. \& Canizares,~C.~R. 1985,
ApJ, 288, 703

\bibitem[Kuiper, Hermsen, \& Mendez(2004)]{Kuiper Hermsen Mendez 2004}
Kuiper,~L., Hermsen,~W. \& Mendez,~M. 2004, ApJ, 613, 1173

\bibitem[Kuiper et al.(2006)]{Kuiper et al. 2006}
Kuiper,~L., Hermsen,~W., den Hartog,~P.~R. \& Collmar,~W. 2006, ApJ, 645, 556

\bibitem[Kuiper, den Hartog, \& Hermsen(2008)]{Kuiper den Hartog Hermsen 2008}
Kuiper,~L., den Hartog,~P.~R. \& Hermsen,~W. 2008, 
Conference Proc. ``3rd International Maxi Workshop,''
10-12 June, 2008, Wako, Saitama, Japan (arXiv:0810.4801)


\bibitem[Kuiper \& Hermsen(2009)]{Kuiper Hermsen 2009}
Kuiper,~L. \& Hermsen,~W. 2009, A\&A, 501, 1031

\bibitem[Marsden \& White(2001)]{Marsden White 2001}
Marsden,~D. \& White,~N.~E. 2001, ApJ, 551, L155


\bibitem[Mereghetti \& Stella(1995)]{Mereghetti Stella 1995}
Mereghetti,~S. \& Stella,~L. 1995, ApJ, 442, L17

\bibitem[Mitsuda et al.(2007)]{Mitsuda et al. 2007}
Mitsuda,~K. et al. 2007, PASJ, 59, S1 

\bibitem[Morii et al.(2003)]{Morii et al. 2003}
Morii,~M., Sato,~R., Kataoka,~J. \& Kawai,~N. 2003,
PASJ, 55, L45

\bibitem[Nakagawa et al.(2009)]{Nakagawa et al. 2009}
Nakagawa,~Y.~E., Yoshida,~A., Yamaoka,~K. \& Shibazaki,~N. 2009, PASJ, 61, 109

\bibitem[{\"O}zel(2001)]{Ozel 2001}
{\"O}zel,~F. 2001, ApJ, 563, 276

\bibitem[Press et al.(1992)]{Press et al.(1992)}
Press,~W.~H. et al. 1992,
``Numerical Recipes in C, Second Edition,'' Cambridge University Press 

\bibitem[Sanbonmatsu \& Helfand(1992)]{Sanbonmatsu Helfand 1992}
Sanbonmatsu,~K.~Y. \& Helfand,~D.~J. 1992, AJ, 104, 2189

\bibitem[Serlemitsos et al.(2007)]{Serlemitsos et al. 2007}
Serlemitsos,~P.~J. 2007, PASJ, 59, S9

\bibitem[Takahashi et al.(2007)]{Takahashi et al. 2007}
Takahashi,~T. et al. 2007, PASJ, 59, S35

\bibitem[Terada et al.(2008)]{Terada et al. 2008}
Terada et al. 2008, PASJ, 60, S25

\bibitem[Thompson \& Beloborodov(2005)]{Thompson Beloborodov 2005}
Thompson,~C. \& Beloborodov,~A.~M. 2005, ApJ, 634, 565

\bibitem[Tian \& Leahy(2007)]{Tian Leahy 2007}
Tian,~W.~W. \& Leahy,~D.~A. 2007, ApJ, 677, 292

\bibitem[van Paradijs, Taam \& van den Heuvel(1995)]{van-Paradijs Taam van-den-Heuvel 1995}
van Paradijs,~J., Taam,~R.~E. \& van den Heuvel,~E.~P.~J. 1995, A\&A, 299, L41

\bibitem[Vasisht \& Gotthelf(1997)]{Vasisht Gotthelf 1997}
Vasisht,~G. \& Gotthelf,~E.~V. 1997, ApJ, 486, L129

\bibitem[Woods \& Thompson(2006)]{Woods Thompson 2006}
Woods,~P.~M. \& Thompson,~C.,
``Soft Gamma repeaters and anomalous X-ray pulsars: magnetar candidates,'' in
Compact Stellar X-ray Sources, edited by Lewin,~W.~H.~G. and van~der~Klis,~M.,
Cambridge University Press, New York, 2006, pp. 547--586 (astro-ph/0406133).

\bibitem[Yamauchi \& Koyama(1993)]{Yamauchi Koyama 1993}
Yamauchi,~S. \& Koyama,~K. 1993, ApJ, 404, 620

\end{thebibliography}
\end{document}